\documentclass[conference]{IEEEtran}
\IEEEoverridecommandlockouts
\usepackage{cite}
\usepackage{amsmath,amssymb,amsfonts}
\usepackage{algorithmic}
\usepackage{graphicx}
\usepackage{textcomp}
\usepackage{xcolor}
\def\BibTeX{{\rm B\kern-.05em{\sc i\kern-.025em b}\kern-.08em
    T\kern-.1667em\lower.7ex\hbox{E}\kern-.125emX}}

\usepackage[ruled, vlined, linesnumbered]{algorithm2e}
\usepackage{soul}
\usepackage{enumitem}
\usepackage{multirow}
\usepackage{booktabs}
\usepackage{array}
\usepackage{pifont}
\usepackage{tikz}
\usepackage{pgfplots}
\usepackage{amsmath}
\usepackage{harpoon}
\usepackage{url}
\usepackage{bm} % for bold math notations
\usepackage{hyperref}
\hypersetup{
    colorlinks=true,
    linkcolor=[RGB]{0, 0, 128},      % dark blue（internal links）
    citecolor=[RGB]{0, 102, 0},      % dark green（citations）
    urlcolor=[RGB]{102, 0, 102}      % purple（URLs）
}

\usepackage[subrefformat=parens, labelformat=parens]{subfig}
\usepackage[font=small]{caption}
\begin{document}

%% COMMENTS
\definecolor{lightgray}{gray}{0.9}
\definecolor{lightblue}{rgb}{0.9,0.9,1}
\definecolor{LightMagenta}{rgb}{1,0.5,1}
\definecolor{red}{rgb}{1,0,0}
\definecolor{brightgreen}{rgb}{0.4, 1.0, 0.0}

\newcommand\couldremove[1]{{\color{lightgray} #1}}
\newcommand{\remove}[1]{}
\newcommand{\move}[2]{ {\textcolor{Purple}{ \bf --- MOVE #1: --- }} {\textcolor{Orchid}{#2}} }

%% FOR COMMENTS & EDITS
\newcommand{\hlc}[2][yellow]{ {\sethlcolor{#1} \hl{#2}} }
\newcommand\note[1]{\hlc[SkyBlue]{-- #1 --}} % highlighted notes of other colors.
% For colors info from xcolor package, check out:
% http://en.wikibooks.org/wiki/LaTeX/Colors

\newcommand\mynote[1]{\hlc[yellow]{#1}}
\newcommand\tingjun[1]{\hlc[yellow]{TC: #1}}
\newcommand\tom[1]{\hlc[brightgreen]{TOM: #1}}
\newcommand\andy[1]{\hlc[pink]{Andy: #1}}
\newcommand\change[1]{{\color{blue} {#1}}}

% Revision highlights
\newcommand{\TODO}[1]{\textcolor{red}{#1}}
\newcommand{\revise}[1]{\textcolor{blue}{#1}}

% \newcommand\mynote[1]{\hlc[yellow]{#1}}
% \newcommand\tingjun[1]{\hlc[yellow]{TC: #1}}
% \newcommand\zhihui[1]{\hlc[LightMagenta]{ZG: #1}}
% \newcommand\tom[1]{\hlc[brightgreen]{TOM: #1}}
% % \newcommand\xiao[1]{\hlc[yellow]{XZ: #1}}
% \newcommand\change[1]{{\color{blue} {#1}}}

\newcommand{\myparatight}[1]{\vspace{0.25ex}\noindent\textbf{#1~}}

\newcommand{\cmark}{\ding{51}}%
\newcommand{\xmark}{\ding{55}}%
\newcommand{\greencheck}{\color[HTML]{3C8031}{\cmark}}
\newcommand{\redcross}{\color[HTML]{ED1B23}{\xmark}}
\newcommand{\greenno}{\color[HTML]{3C8031}{\textbf{No}}}
\newcommand{\redyes}{\color[HTML]{ED1B23}{\textbf{Yes}}}
\newcommand{\greenlow}{\color[HTML]{3C8031}{\textbf{Low}}}
\newcommand{\redhigh}{\color[HTML]{ED1B23}{\textbf{High}}}

\newcommand*\circledwhite[1]{\tikz[baseline=(char.base)]{
            \node[shape=circle,draw,inner sep=0.6pt] (char) {\scriptsize{#1}};}}

\newcommand*\circled[1]{\tikz[baseline=(char.base)]{
            \node[shape=circle,draw,fill=black,text=white,inner sep=1pt] (char) {\scriptsize{#1}};}}

%% paper-specific variable/notation
\newcommand{\name}{{RISE}}
\newcommand{\namebf}{{\textbf{RISE}}}

\newcommand{\agora}{Agora}
\newcommand{\agorabf}{\textbf{Agora}}

\newcommand{\armavec}{\sf Savannah-mc (arma-vec)}
\newcommand{\armacube}{\sf Savannah-mc (arma-cube)}

%%This is a special cell that allows using \\ inside the cell to add a new line
\newcommand{\specialcell}[2][c]{%
  \begin{tabular}[#1]{@{}c@{}}#2\end{tabular}}

\newcommand{\iu}{{j}}

\newcommand{\littlesum}{\mathop{\textstyle\sum}}
\newcommand{\littleint}{\mathop{\textstyle\int}}

% config
\newcommand{\siso}{SISO}
\newcommand{\mimoTwoByTwo}{2$\times$2}
\newcommand{\mimoFourByFour}{4$\times$4}

% lib
\newcommand{\bbdev}{\textsf{bbdev}}

% PHY
\newcommand{\scs}{{\Delta f}}
\newcommand{\scNum}{N_{\textrm{sc}}}
\newcommand{\sampRate}{F_{\textrm{s}}}
\newcommand{\fftSize}{N_{\textrm{FFT}}}
\newcommand{\chMat}{{\textbf{H}}}
\newcommand{\chVec}{{\textbf{h}}}
\newcommand{\precodeMat}{{\textbf{P}}}

% units
\newcommand{\usec}{$\mu$s} % us or usec
\newcommand{\msec}{ms}       % ms or msec

% DSP stages
\newcommand{\fft}{\textsf{fft}}
\newcommand{\ifft}{\textsf{ifft}}
\newcommand{\csi}{\textsf{csi}}
\newcommand{\precode}{\textsf{precode}}
\newcommand{\encode}{\textsf{enc}}
\newcommand{\decode}{\textsf{dec}}
\newcommand{\modul}{\textsf{modul}}
\newcommand{\demod}{\textsf{demod}}
\newcommand{\equal}{\textsf{equal}}

% LDPC parameters
\newcommand{\tbSize}{T}
\newcommand{\tbCrcSize}{T_{\textrm{crc}}}
\newcommand{\cbSize}{K_{\textrm{cb}}}
\newcommand{\cbNum}{N_{\textrm{cb}}}
\newcommand{\liftingSize}{Z_{c}}
\newcommand{\liftingSizeSet}{\mathbf{\Theta}}
\newcommand{\fillerBitNum}{N_{\textrm{filler}}}

\newcommand{\codeRate}{R}

\newcommand{\throughput}{Tp}
\newcommand{\codingTime}{t}
\newcommand{\informationBits}{K'}

% Thresholds
\newcommand{\thresPower}{\theta_{\textrm{Power}}}
\newcommand{\thresOtsu}{\theta_{\textrm{Otsu}}}
\newcommand{\thresPSD}{\theta_{\textrm{PSD}}}
\newcommand{\thresIoU}{\theta_{\textrm{IoU}}}

% Real-time parameters
\newcommand{\tput}{\lambda} % throughput
\newcommand{\lat}{l} % latency

% MO parameters
\newcommand{\se}{E}

\newcommand{\myAbs}[1]{\left|{#1}\right|}
\newcommand{\myAng}[1]{\angle{#1}}
\newcommand{\myConjugate}[1]{{#1}^{*}}
\newcommand{\myTranspose}[1]{{#1}^{\top}}
\newcommand{\myHermitian}[1]{{#1}^{H}}
\newcommand{\myIsFunc}[1]{\mathbf{1}\{#1\}}

% Angle
\newcommand{\AoD}{\phi}
\newcommand{\AoDVec}{\bm{\upphi}}
\newcommand{\AoDbf}{\boldsymbol\phi}
\newcommand{\AoDDirectional}{\Phi}
\newcommand{\az}{\phi}
\newcommand{\azVec}{\bm{\upphi}}
\newcommand{\azVecUE}{\bm{\upphi}_{\textrm{UE}}}
\newcommand{\azbf}{\boldsymbol\phi}
\newcommand{\el}{\psi}
\newcommand{\elbf}{\boldsymbol\psi}

% Element Setup
\newcommand{\ElemComp}{w}
\newcommand{\ElemCompbf}{\mathbf{w}}
\newcommand{\ElemCompNew}{w^\prime}
\newcommand{\ElemCompNewbf}{\mathbf{w}^\prime}
\newcommand{\ElemAmp}{A}
\newcommand{\ElemAmpbf}{\mathbf{A}}
\newcommand{\ElemPhase}{\theta}
\newcommand{\ElemPhasebf}{\boldsymbol\theta}
\newcommand{\steer}{s}
\newcommand{\steerVec}{\mathbf{s}}
\newcommand{\steermat}{\mathbf{S}}
\newcommand{\beamPattern}{BP}

\newcommand{\bw}{B}
\newcommand{\carrierFreq}{f_{c}}
\newcommand{\carrierWave}{\lambda}

\newcommand{\csiMat}{\mathbf{H}}

\newcommand{\ASA}[2]{\textrm{ASA}({#1},{#2})}
\newcommand{\antNum}{N}
\newcommand{\antIdx}{n}
\newcommand{\antDist}{d}
\newcommand{\subarrayNum}{M}
\newcommand{\subarraySet}{\mathcal{M}}
\newcommand{\subarrayIdx}{m}
\newcommand{\subarrayAntNum}{N_{s}}
\newcommand{\subarrayAntIdx}{n}
\newcommand{\subarrayAntDist}{d}

\newcommand{\setSubarray}{\mathcal{A}}
\newcommand{\subarrayAlloc}{a}
\newcommand{\subarrayAllocVec}{\mathbf{a}}
\newcommand{\subarrayAllocMat}{\mathbf{A}}
\newcommand{\subarrayAllocSet}{\mathbb{A}}

% beamforming
\newcommand{\bfWeight}{w}
\newcommand{\bfWeightVec}{\mathbf{w}}
\newcommand{\bfAmp}{A}
\newcommand{\bfAmpVec}{\mathbf{A}}
\newcommand{\bfPhase}{\theta}
\newcommand{\bfPhaseVec}{\boldsymbol{\theta}}
\newcommand{\bfGain}{g}
\newcommand{\bfGainSig}[1]{g^{\textrm{sig}}_{#1}}
\newcommand{\bfGainInt}[2]{g^{\textrm{int}}_{{#1}\rightarrow{#2}}}

\newcommand{\power}{\mathcal{P}}
\newcommand{\powerSignal}{\mathcal{S}}
\newcommand{\powerSignalDiff}{d\mathcal{S}}
\newcommand{\powerInterf}{\mathcal{I}}
\newcommand{\powerInterfDiff}{d\mathcal{I}}
\newcommand{\powerNoise}{N}

% network model
\newcommand{\userNum}{U}
\newcommand{\userIdx}{u}
\newcommand{\userSet}{\mathcal{U}}

\newcommand{\userNumSub}{K}

\newcommand{\userSelected}{k}
\newcommand{\userSelectedNum}{K}
\newcommand{\userSelectedSet}{\mathcal{K}}

\newcommand{\userAngle}{\phi}
\newcommand{\userWeight}{\alpha}

\newcommand{\baseSNR}{\gamma}
\newcommand{\SNR}{\mathsf{SNR}}
\newcommand{\SNRMax}{\mathsf{SNR}^{\textrm{max}}}
\newcommand{\SINR}{\mathsf{SINR}}
\newcommand{\SINRMax}{\mathsf{SINR}^{\textrm{max}}}
\newcommand{\Capacity}{T}
\newcommand{\Rate}{R}
\newcommand{\RateMax}{\Rate^{\textrm{max}}}
\newcommand{\RateAvg}{\widebar{\Rate}}
\newcommand{\CapacityMax}{\Tilde{T}}
\newcommand{\suppress}{\alpha}

% Proportional Fair
\newcommand{\RateHist}{\widebar{\Rate}}

\newcommand{\past}{p}
\newcommand{\decay}{\beta}

\newcommand{\RateMean}{\Bar{R}}
\newcommand{\JFI}{\mathsf{JFI}}

%% Bold rows in a table
\newcolumntype{+}{>{\global\let\currentrowstyle\relax}}
\newcolumntype{^}{>{\currentrowstyle}}
\newcommand{\rowstyle}[1]{%
  \gdef\currentrowstyle{#1}%
  #1\ignorespaces
}

% Matrix representation
\newenvironment{spmatrix}[1]
 {\def\mysubscript{#1}\mathop\bgroup\begin{bmatrix}}
 {\end{bmatrix}\egroup_{\textstyle\mathstrut\mysubscript}}

\newcommand{\bigO}{\mathcal{O}} % Open at top left
\newcommand{\conv}{\ast}

\title{RISE: Real-time Image Processing for Spectral Energy Detection and Localization}

\author{
\IEEEauthorblockN{Chung-Hsuan Tung, Zhenzhou Qi, Tingjun Chen}
\IEEEauthorblockA{
Department of Electrical and Computer Engineering, Duke University \\
\{chunghsuan.tung, zhenzhou.qi, tingjun.chen\}@duke.edu}
}

\maketitle

\begin{abstract}
Energy detection is widely used for spectrum sensing, but accurately localizing the time and frequency occupation of signals in real-time for efficient spectrum sharing remains challenging.
To address this challenge, we present {\name}, a software-based spectrum sensing system designed for real-time signal detection and localization.
{\name} treats time-frequency spectrum plots as images and applies adaptive thresholding, morphological operations, and connected component labeling with a multi-threaded architecture.
We evaluate {\name} using both synthetic data and controlled over-the-air (OTA) experiments across diverse signal types.
Results show that {\name} satisfies real-time latency constraints while achieving a probability of detection of {80.42\%} at an intersection-over-union (IoU) threshold of 0.4.
{\name} sustains a raw I/Q input rate of {3.2}\thinspace{Gbps} for {100}\thinspace{MHz} bandwidth sensing with time and frequency resolutions of {10.24}\thinspace{\usec} and {97.6}\thinspace{kHz}, respectively.
Compared to Searchlight, a representative energy-based method, {\name} achieves 20.51$\times$ lower latency and 22.31\% higher IoU.
Compared to machine learning baselines, {\name} improves IoU by 56.02\% over DeepRadar while meeting the real-time deadline, which a GPU-accelerated U-Net exceeds by 213.38$\times$.
\end{abstract}


\section{Introduction}
\label{sec: intro}

%%%%%%%%%%%%%%%%%%%%%%%%%%%%%%%%%%%%%%%%%%%%%%%%%%%%%%%%%%%%%%%%%%%%%%%%%%%%%%%%
%%%%%%%%%%%%%%%%%%%%%%%%%%%%%%%%%%%%%%%%%%%%%%%%%%%%%%%%%%%%%%%%%%%%%%%%%%%%%%%%

%%%%%%%%%%%%%%%%%%%%%%%%%%%%%%%%%%%%%%%%%%%%%%%%%%%%%%%%%%%%%%%%%%%%%%%%%%%%%%%%
% introduce the need for spectrum sensing.
Spectrum sensing is a key enabler of efficient spectrum sharing in the 5G-and-beyond era, given the limited resources and proliferating wireless applications~\cite{sarkar2021deepradar, reus2023senseoran, nika2014towards, singh2015coordination, yuksel2016pervasive, qi2023programmable, guimaraes2022spectrum, zheleva2023radio}.
Machine learning (ML)-based solutions often use object detection, such as the YOLO model~\cite{redmon2016you, sarkar2021deepradar, soltani2022finding, sarkar2024radyololet, rahman2024speclearn, reus2023senseoran, reihan2024low}, or semantic segmentation~\cite{uvaydov2024stitching, subedi2024seek, huynh2024srnet}.
Those data-driven solutions often require prepared datasets and have limited generalizability to unseen patterns, as wireless signals exhibit highly variable patterns.
% The model may learn the dimensions of an object, but wireless signals can vary in length and occupied spectrum, e.g., short spikes for control message exchange are common in wireless communication but an extreme case in object detection.
%
As the spectrum becomes more congested when serving versatile protocols~\cite{scisrs}, energy-based detection can identify signals without prior knowledge~\cite{searchlight_slides}.
Though energy detection is often considered a fast and simple solution, challenges exist for obtaining good time/frequency resolution, adaptive noise floor estimation, and accommodating various signal patterns.
Beyond the detection of signal presence, localizing their time/frequency occupation is necessary for the downstream application to interpret the signal or avoid resource conflict.

%%%%%%%%%%%%%%%%%%%%%%%%%%%%%%%%%%%%%%%%%%%%%%%%%%%%%%%%%%%%%%%%%%%%%%%%%%%%%%%%
% process of energy detection
The energy detection-based localization of signals generally includes three steps:
(\emph{i}) obtain the time-frequency (TF) plot,
(\emph{ii}) estimate the noise floor to identify regions of interest, and
(\emph{iii}) localize the precise range of the energy blocks.
As the localization requires both time and frequency occupation, the system performs Fourier Transform operations to convert a chunk of I/Q samples received by the radio, or spectrum sensor, into the frequency domain.
This conversion determines the fundamental time and frequency resolution $\Delta t = N_{\text{FFT}}/f_s$ and $\Delta f = f_s/N_{\text{FFT}}$: one power spectral density (PSD) corresponds to I/Q samples in a time duration of $\Delta t$ with a sampling rate $f_s$ generated via a Fast Fourier Transform (FFT) with $N_{\text{FFT}}$ points.
Multiple such PSDs can then be stacked to create a two-dimensional (2D) TF plot, or spectrogram.
%%%%%%%%%%%%%%%%%%%%%%%%%%%%%%%%%%%%%%%%%%%%%%%%%%%%%%%%%%%%%%%%%%%%%%%%%%%%%%%%
% mention baseline
Representative energy detection-based systems, such as Searchlight~\cite{bell2023searchlight}, demonstrate that accurate localization is possible, but achieving this capability under real-time constraints remains challenging.

%%%%%%%%%%%%%%%%%%%%%%%%%%%%%%%%%%%%%%%%%%%%%%%%%%%%%%%%%%%%%%%%%%%%%%%%%%%%%%%%
% explain real-time requirement and multi-threaded model
Spectrum sensing requires a high-throughput, uninterrupted system with lossless input stream handling.
This requirement differs from that of wireless communication, which relies on protocol-level redundancy to guarantee transmission quality.
For example, a 5G cellular base station using a time-division duplexing (TDD) mode defined by 3GPP~\cite{3gpp38.213, 3gpp38.101-2} allows the uplink to occupy only 23\% of the transmission, while the cyclic prefix (CP) as the guard interval and the guard band subcarrier account for 6.7\% and 22.7\% redundancy, respectively~\cite{qi2024savannah}.
However, a sensing system without prior knowledge is expected not to miss any of the received samples, ensuring complete coverage over the observation time and frequency range.
While the processing deadline is clear for a communication system, it can vary for a sensing system depending on downstream applications and the desired time window size.
Previous baseband processing works ~\cite{ding2020agora, qi2024savannah, wang2024understanding} implemented real-time software to meet high throughput and low latency; however, their timing assumptions do not directly extend to spectrum sensing workloads.
While the idea of multi-threaded architecture is transferable, the data arrangement must be carefully designed for the spectrum sensing system.

To address these challenges, we present {\name}---\underline{R}eal-time \underline{I}mage processing for \underline{S}pectral \underline{E}nergy detection and localization.
Implemented in C++, {\name} provides real-time spectrum awareness by completing the processing of each TF plot within its generation period.
Specifically, we formulate energy-based spectrum sensing as an image processing pipeline operating on TF plots, consisting of
(\emph{i}) PSD-based noise floor estimation,
(\emph{ii}) binarization with adaptive Otsu thresholding~\cite{otsu1975threshold},
(\emph{iii}) morphological operations (MOs) for energy block consolidation, and
(\emph{iv}) connected component labeling (CCL) for bounding boxes localization.
Compared to existing works such as Searchlight~\cite{bell2023searchlight}, {\name} eliminates exhaustive convolution kernel search and manual threshold tuning, reducing the algorithmic complexity from $\bigO(T^{2}F^{2})$ to $\bigO(TF\log F)$ for each TF plot with dimension $(T,F)$.

%%%%%%%%%%%%%%%%%%%%%%%%%%%%%%%%%%%%%%%%%%%%%%%%%%%%%%%%%%%%%%%%%%%%%%%%%%%%%%%%

We evaluate {\name} using synthetic signals under simulation and over-the-air (OTA) settings with USRP X310 software-defined radios (SDRs).
{\name} meets the real-time processing requirement for {100}\thinspace{MHz} wideband sensing ({100}\thinspace{MSps} throughput) using 16 CPU cores and a multi-threaded architecture with ping-pong buffering, while achieving an average probability of detection of 80.42\% at an intersection-over-union (IoU) threshold of 0.4 under {14}\thinspace{dB} SNR.
Compared to Searchlight~\cite{bell2023searchlight}, {\name} reduces the processing latency by {20.51}$\times$ on average at the algorithm level.
For the ML-based object detection, {\name} increases the average IoU by 56.02\% while reducing the processing latency by {1.65}$\times$ compared to DeepRadar~\cite{sarkar2021deepradar}.
For ML-based image segmentation, {\name} meets real-time deadlines under a multi-core CPU configuration, which U-Net~\cite{uvaydov2024stitching} exceeds by {213.38}$\times$ even with GPU acceleration, despite achieving competitive detection performance.

%%%%%%%%%%%%%%%%%%%%%%%%%%%%%%%%%%%%%%%%%%%%%%%%%%%%%%%%%%%%%%%%%%%%%%%%%%%%%%%%
% summarize the contribution
The contributions of {\name} are summarized as follows:
\begin{itemize}[leftmargin=*, topsep=2pt, itemsep=1pt]
\item We propose an efficient energy detection and localization algorithm for spectrum sensing, achieving high detection accuracy and low complexity using lightweight image processing primitives, including adaptive binarization, morphological operations, and connected component labeling;
\item We implement the proposed algorithm as a real-time sensing software system in C++ using a multi-threaded architecture that supports high-throughput, low-latency spectrum data processing without interruptions;
\item We formulate evaluation metrics for energy localization and conduct experiments on both synthetic datasets and real-world measurements to analyze the system performance with respect to various design parameters.
\end{itemize}
\noindent
We open-source {\name} at \url{https://github.com/functions-lab/rise}.

\section{Related Work}
\label{sec: related_work}

%%%%%%%%%%%%%%%%%%%%%%%%%%%%%%%%%%%%%%%%%%%%%%%%%%%%%%%%%%%%%%%%%%%%%%%%%%%%%%%%
\myparatight{Spectrum sensing systems.}
Spectrum sensing is fundamental to dynamic spectrum access and sharing, in which users must detect the presence and occupation of coexisting transmissions over the wireless channel.
Representative applications include co-existence of Wi-Fi, Bluetooth, and LTE in the {2.4}\thinspace{GHz} band, radar detection in the CBRS ({3.5}\thinspace{GHz}) band~\cite{sarkar2021deepradar, reus2023senseoran}, and military use~\cite{bell2023searchlight}.
Spectrum sensing techniques~\cite{arjoune2019comprehensive} include energy detection~\cite{bell2023searchlight}, cyclo-stationary analysis~\cite{guddeti2019sweepsense}, matched filtering, covariance-based methods, and machine learning (ML)~\cite{sarkar2021deepradar, reus2023senseoran}.
While many studies focus on signal existence, fewer address time--frequency localization.
A common design choice is to assume steady-state signals over partitioned intervals~\cite{sarkar2021deepradar, reus2023senseoran}, relying on prior knowledge of the target signals.
Learning-based methods similarly depend on known signal features.
In contrast, energy-based approaches require no prior knowledge and localize transient signals by identifying contiguous energy blocks on TF plots~\cite{bell2023searchlight}.

%%%%%%%%%%%%%%%%%%%%%%%%%%%%%%%%%%%%%%%%%%%%%%%%%%%%%%%%%%%%%%%%%%%%%%%%%%%%%%%%
\myparatight{Real-time spectrum sensing.}
Most existing spectrum sensing systems rely on hardware acceleration or still fall short of real-time processing requirements.
For example, SenseORAN~\cite{reus2023senseoran} implements a YOLO-based detector~\cite{redmon2016you} as an xApp in the Near-RT RIC of O-RAN to sense CBRS radar signals with a single CPU core, where processing a {10}\thinspace{\msec} spectrogram sampled at $f_s={15.36}\thinspace\textrm{MSps}$ requires {866}\thinspace{\msec}.
DeepSense~\cite{uvaydov2021deepsense} deploys a CNN model on an Xilinx Zynq-7000 FPGA, but processes recorded Wi-Fi and LTE signals offline.
DeepRadar~\cite{sarkar2021deepradar} similarly applies YOLO-based detection for CBRS radar signals on an AIR-T SDR with Nvidia Jetson TX2 GPUs, achieving a throughput of only {2}\thinspace{MSps}.

\section{Problem Formulation}
\label{sec: prelim}

%%%%%%%%%%%%%%%%%%%%%%%%%%%%%%%%%%%%%%%%%%%%%%%%%%%%%%%%%%%%%%%%%%%%%%%%%%%%%%%%
% System Design Image for the next section (brought forward)
%%%%%%%%%%%%%%%%%%%%%%%%%%%%%%%%%%%%%%%%%%%%%%%%%%%%%%%%%%%%%%%%%%%%%%%%%%%%%%%%
%% figure begins
\begin{figure*}[!t]
    \centering
    \includegraphics[width=0.99\textwidth]{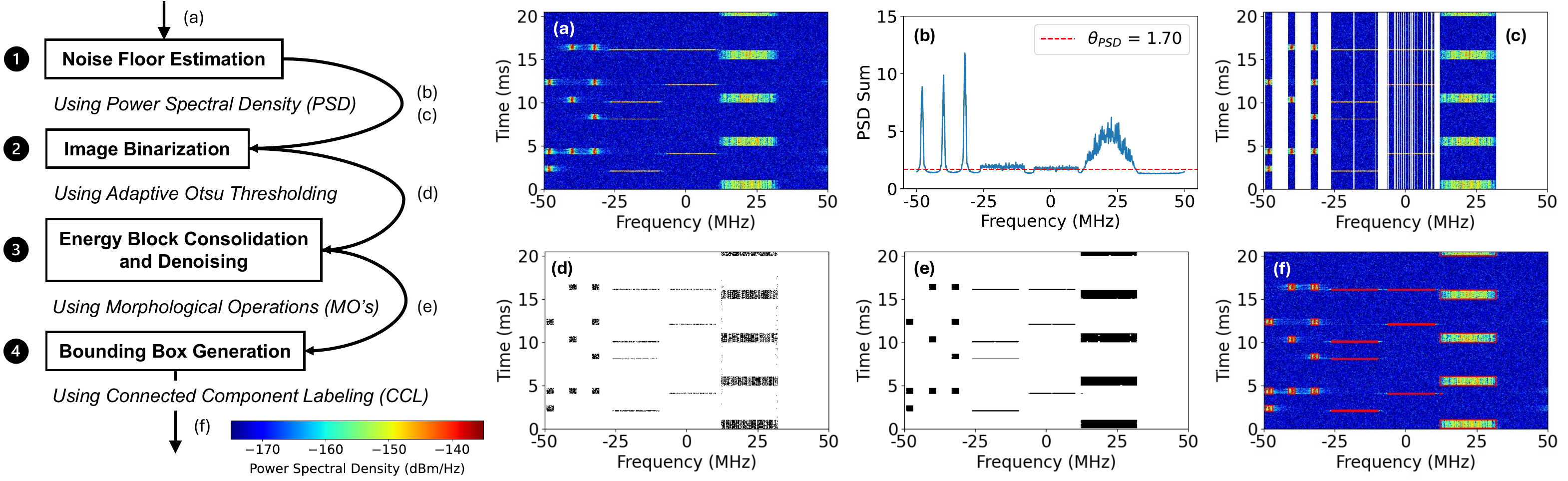}
    \caption{
    Overview of the image-processing pipeline and intermediate representations for signal detection and localization in {\name}.}
    \label{fig: sys_flow}
    \vspace{-3mm}
\end{figure*}
%% figure ends
%%%%%%%%%%%%%%%%%%%%%%%%%%%%%%%%%%%%%%%%%%%%%%%%%%%%%%%%%%%%%%%%%%%%%%%%%%%%%%%%
%%%%%%%%%%%%%%%%%%%%%%%%%%%%%%%%%%%%%%%%%%%%%%%%%%%%%%%%%%%%%%%%%%%%%%%%%%%%%%%%

This section formulates real-time energy-based spectrum sensing as a joint detection and time--frequency localization problem under bounded latency constraints.

%%%%%%%%%%%%%%%%%%%%%%%%%%%%%%%%%%%%%%%%%%%%%%%%%%%%%%%%%%%%%%%%%%%%%%%%%%%%%%%%
%%%%%%%%%%%%%%%%%%%%%%%%%%%%%%%%%%%%%%%%%%%%%%%%%%%%%%%%%%%%%%%%%%%%%%%%%%%%%%%%
\subsection{Energy-based Spectrum Sensing}

%%%%%%%%%%%%%%%%%%%%%%%%%%%%%%%%%%%%%%%%%%%%%%%%%%%%%%%%%%%%%%%%%%%%%%%%%%%%%%%%
\subsubsection{Detection and Localization Formulation}
\label{sub2sec: prob_formulation}

We consider an uninterrupted stream of in-phase and quadrature (I/Q) samples.
The receiver converts I/Q samples into a time-frequency (TF) plot by applying a non-overlapping $N_{\text{FFT}}$-point FFT over consecutive segments.
At a sampling rate $f_s$, the time and frequency resolutions are $\Delta t = N_{\text{FFT}}/f_s$ and $\Delta f = f_s/N_{\text{FFT}}$, respectively.
Stacking $T$ segments yields a TF plot of size $T \times F$, where $F=N_{\text{FFT}}$.

A signal occupying time interval $[t_0,t_1]$ and frequency band $[f_0,f_1]$ corresponds to a high-intensity region on the TF plot.
We represent each signal by a bounding box $B$; $N_{gt}$ ground-truth boxes $\mathcal{B}$ and $N_d$ detected boxes $\hat{\mathcal{B}}$ are written by
\begin{equation}
\label{eq: box_list_gt_detect}
B=(f_0,f_1,t_0,t_1), \quad
\mathcal{B} = \{B_i\}_{i=1}^{N_{gt}}, \quad
\hat{\mathcal{B}} = \{\hat{B}_j\}_{j=1}^{N_{d}}.
\end{equation}

%%%%%%%%%%%%%%%%%%%%%%%%%%%%%%%%%%%%%%%%%%%%%%%%%%%%%%%%%%%%%%%%%%%%%%%%%%%%%%%%

\subsubsection{Challenges in Signal Localization}
\label{sub2sec: det_loc_challenges}

Energy-based signal localization aims to identify high-intensity regions and infer their time--frequency extent.
Both tasks rely solely on signal magnitude, as energy measured in the frequency domain is equivalent to that in the time domain up to a constant scaling.
While this approach assumes a discernible energy rise over the noise floor (e.g., SNR $>$ {0}\thinspace{dB}), localizing a signal reduces to detecting contiguous high-energy regions on the TF plot.

\myparatight{Identifying high-intensity areas.}
A common approach is to classify TF bins whose power exceeds a threshold as signal.
This requires selecting both an energy threshold (noise floor estimation) and an effective detector size.
Noise floor estimation often depends on environment-specific knowledge~\cite{bell2023searchlight},
while selecting the detector size may require exhaustive search~\cite{bell2023searchlight} or assumptions on signal duration~\cite{sarkar2021deepradar}.

\myparatight{Inferring signal time--frequency occupation.}
Even after high-intensity regions are identified, precise boundary inference remains challenging due to non-uniform power distributions.
Signals may exhibit sharp or gradual boundary transitions, causing local power change-based methods to fragment detections or blur signal edges, especially in the presence of internal fluctuations and noise peaks.
Prior work such as Searchlight~\cite{bell2023searchlight} applies upfront averaging to smooth power variations, but this trades time--frequency resolution for robustness and may obscure signal boundaries.

%%%%%%%%%%%%%%%%%%%%%%%%%%%%%%%%%%%%%%%%%%%%%%%%%%%%%%%%%%%%%%%%%%%%%%%%%%%%%%%%
%%%%%%%%%%%%%%%%%%%%%%%%%%%%%%%%%%%%%%%%%%%%%%%%%%%%%%%%%%%%%%%%%%%%%%%%%%%%%%%%
\subsubsection{Detection and Localization Metrics}
\label{sub2sec: detect_metric}

We evaluate spectrum sensing performance in terms of both signal detection and localization quality.
While detection concerns whether a signal is identified, localization further requires estimating its time--frequency occupation.
To jointly capture both aspects, we define the probability of detection $P_d$ and the false alarm rate $P_{fa}$ using decision rules based on bounding-box overlap.

The overlap between a ground-truth box $B$ and a detected box $\hat{B}$ in \eqref{eq: box_list_gt_detect} is measured by the intersection-over-union (IoU), where
$\text{IoU}(B, \hat{B}) = ({B \cap \hat{B}})/({B \cup \hat{B}})$.
We compute the IoU matrix $\mathbf{I} = [I_{ij}] \in [0,1]^{N_{gt} \times N_d}$, for pair-wise IoU $I_{ij} = \text{IoU}(B_i, \hat{B}_j)$.
Given an IoU threshold $\thresIoU$, a ground-truth signal is considered successfully detected if it overlaps with at least one detected box above the threshold.
Accordingly, the numbers of true detections $N_t$ and false detections $N_f$ are defined using the indicator function $\mathbf{1}\{\cdot\}$:
\begin{equation}\label{eq_td_fd_by_thres_iou}
N_t = \littlesum_{i=1}^{N_{gt}} \mathbf{1}\!\left\{ \max_j I_{ij} > \thresIoU \right\},
N_f = \littlesum_{j=1}^{N_d} \mathbf{1}\!\left\{ \max_i I_{ij} < \thresIoU \right\}.
\end{equation}
$P_d$ and $P_{fa}$ are then defined as
\begin{equation}\label{eq_pd_pfa}
P_d \triangleq {N_t}/{N_{gt}}, \quad
P_{fa} \triangleq {N_f}/{N_d}.
\end{equation}

%%%%%%%%%%%%%%%%%%%%%%%%%%%%%%%%%%%%%%%%%%%%%%%%%%%%%%%%%%%%%%%%%%%%%%%%%%%%%%%%
%%%%%%%%%%%%%%%%%%%%%%%%%%%%%%%%%%%%%%%%%%%%%%%%%%%%%%%%%%%%%%%%%%%%%%%%%%%%%%%%
\subsection{Real-time Processing Requirements}
\label{subsec: rt_challenge}

Two widely used metrics for characterizing system speed are throughput ($\tput$) and latency ($\lat$).
Throughput measures the amount of data processed per unit time; latency measures the end-to-end time of a task.
They are related but not equivalent.

{\name} targets real-time spectrum sensing, which requires both high throughput and bounded latency.
High throughput is necessary for wideband sensing, as sampling rates $f_s$ directly translate into the raw I/Q input rate (e.g., {100}\thinspace{MHz} sensing with $f_s={100}\thinspace\textrm{MSps}$ corresponds to {3.2}\thinspace{Gbps} with 16-bit I/Q samples).
Meanwhile, spectrum-sharing applications often require timely responses, e.g., to obtain immediate access to wireless physical resources, placing strict latency constraints.

We therefore define real-time spectrum sensing as \textit{bounded-latency processing}, where each task must complete before the next one becomes available, ensuring uninterrupted sensing without backlog accumulation.
To account for runtime variability, we further require that \emph{at least 99\% of tasks satisfy this latency bound}.
Under this definition, meeting the latency requirement is sufficient to sustain the input throughput, whereas high throughput alone does not guarantee real-time operation.

%%%%%%%%%%%%%%%%%%%%%%%%%%%%%%%%%%%%%%%%%%%%%%%%%%%%%%%%%%%%%%%%%%%%%%%%%%%%%%%%
%%%%%%%%%%%%%%%%%%%%%%%%%%%%%%%%%%%%%%%%%%%%%%%%%%%%%%%%%%%%%%%%%%%%%%%%%%%%%%%%
\subsection{Limitations of Existing Approaches}\label{subsec: searchlight_flow}

Most energy-based localization methods share a similar workflow and face challenges of meeting real-time constraints.

%%%%%%%%%%%%%%%%%%%%%%%%%%%%%%%%%%%%%%%%%%%%%%%%%%%%%%%%%%%%%%%%%%%%%%%%%%%%%%%%
% compare to baseline
We use Searchlight~\cite{bell2023searchlight} as a recent, representative example.
Searchlight estimates the noise floor via empirical correction, smooths the TF plot through upfront averaging, and applies thresholded convolution with an exhaustive search over kernel dimensions to detect energy blocks of varying sizes.
Detected block boundaries are then inferred by expanding outward until significant power-rate changes are observed.
While effective in localization accuracy, this design is sensitive to multiple manually tuned parameters, including noise-floor correction and thresholds for convolution and boundary inference.
In addition, exhaustive kernel search incurs high computational overhead, and pruning the search space is difficult without prior knowledge of signal dimensions, complicating the tradeoff between processing latency under real-time throughput constraints and reliable detection across unknown signal types.

Learning-based approaches~\cite{sarkar2021deepradar, uvaydov2024stitching} similarly rely on prior knowledge in the form of training data and typically incur higher computational costs, further limiting their applicability to flexible, real-time spectrum sensing.

These limitations motivate a lightweight, parameter-free energy-localization approach that maintains real-time throughput without assuming prior knowledge of signal patterns.

\section{System Design}
\label{sec: design}

To support real-time spectrum sensing, {\name} integrates efficient and effective image processing algorithms to detect energy blocks, along with a streaming and computation software framework that leverages multi-threaded processing.

%%%%%%%%%%%%%%%%%%%%%%%%%%%%%%%%%%%%%%%%%%%%%%%%%%%%%%%%%%%%%%%%%%%%%%%%%%%%%%%%
\subsection{System Overview}
{\name} localizes signals in the spectrum via energy detection by producing bounding boxes for energy blocks on a TF plot, as illustrated in Fig.~\ref{fig: sys_flow}.
Starting from Fig.~\ref{fig: sys_flow}(a), an example input TF plot, the processing pipeline contains four modules:
\begin{enumerate}[leftmargin=*, topsep=2pt, itemsep=1pt]
    \item[\circled{1}]
    \textbf{Noise floor estimation.} The TF plot is projected along the time axis to estimate the power spectral density (PSD) in Fig.~\ref{fig: sys_flow}(b), which is used to identify frequency components that potentially contain signal energy, as shown in Fig.~\ref{fig: sys_flow}(c).
    \item[\circled{2}]
    \textbf{Image binarization.} For each selected frequency component, adaptive Otsu thresholding~\cite{otsu1975threshold} is applied column-wise to separate signal energy from local noise, binarizing the TF plot, as shown in Fig.~\ref{fig: sys_flow}(d).
    \item[\circled{3}]
    \textbf{Energy block consolidation and denoising.} Morphological operations (MOs) connect fragmented signal regions and suppress isolated noise peaks, producing Fig.~\ref{fig: sys_flow}(e).
    \item[\circled{4}]
    \textbf{Bounding box generation.} Each connected energy block is identified, and its time--frequency range is extracted to generate final bounding boxes in Fig.~\ref{fig: sys_flow}(f).
\end{enumerate}
The detailed operations are elaborated in Sec.~\ref{subsec: imag_proc}.

%%%%%%%%%%%%%%%%%%%%%%%%%%%%%%%%%%%%%%%%%%%%%%%%%%%%%%%%%%%%%%%%%%%%%%%%%%%%%%%%
%%%%%%%%%%%%%%%%%%%%%%%%%%%%%%%%%%%%%%%%%%%%%%%%%%%%%%%%%%%%%%%%%%%%%%%%%%%%%%%%
\subsection{Image Processing Algorithms}
\label{subsec: imag_proc}

%%%%%%%%%%%%%%%%%%%%%%%%%%%%%%%%%%%%%%%%%%%%%%%%%%%%%%%%%%%%%%%%%%%%%%%%%%%%%%%%
\myparatight{Noise floor estimation with PSD.}
To reduce unnecessary computation under wideband sensing, {\name} first performs a coarse frequency-domain pruning to identify regions of interest.
Specifically, the power spectral density (PSD) is estimated by aggregating the TF plot along the time axis, yielding a per-frequency power profile.
Frequency components with low aggregated power are unlikely to contain signals and can be excluded from subsequent processing.
To determine the noise floor, the PSD curve is smoothed using a Savitzky--Golay filter, and the lowest local minimum is selected as the noise floor $N_0$.
A global threshold $\thresPSD$ is then derived from $N_0$.
Frequency columns whose aggregated power falls below $\thresPSD$ are pruned and skipped by adaptive binarization.

%%%%%%%%%%%%%%%%%%%%%%%%%%%%%%%%%%%%%%%%%%%%%%%%%%%%%%%%%%%%%%%%%%%%%%%%%%%%%%%%
\myparatight{Adaptive image binarization using Otsu thresholding.}
After coarse frequency pruning, {\name} applies adaptive binarization to separate signal energy from noise-dominated samples.
A single global threshold is insufficient because noise and signal strength vary spatially across the TF plot.
We therefore perform adaptive thresholding independently for each frequency column.
Otsu thresholding~\cite{otsu1975threshold} is applied to each retained column to select a threshold $\thresOtsu$ that maximizes the between-class variance $\sigma_b^2(\theta)$,
\begin{equation}
\thresOtsu = \arg\max_{\theta} \; \sigma_b^2(\theta),
\end{equation}
yielding a parameter-free binarization that adapts to local signal and noise statistics.
This approach exploits the separation between signal- and noise-dominated intensity distributions without manual threshold tuning.
Pixels above $\thresOtsu$ are classified as foreground, producing a binary TF mask that preserves local signal structure while suppressing background noise.

%%%%%%%%%%%%%%%%%%%%%%%%%%%%%%%%%%%%%%%%%%%%%%%%%%%%%%%%%%%%%%%%%%%%%%%%%%%%%%%%
\myparatight{Energy block consolidation and denoising via MOs.}
After adaptive binarization, foreground pixels form fragmented or irregular clusters due to variations in signal power and residual noise.
{\name} therefore applies morphological operations (MOs) on the binary TF mask to restore spatial connectivity and suppress spatially isolated noise~\cite{alammar2023enhanced}.
{\name} uses opening and closing operations with small, fixed structuring elements (SEs) to remove spurious pixels and fill internal gaps within signal regions.
Opening ($\circ$) removes small disconnected components, and closing ($\bullet$) bridges narrow gaps inside an energy block:
\begin{equation}
A \circ \se = (A \ominus \se) \oplus \se,
\quad
A \bullet \se = (A \oplus \se) \ominus \se.
\end{equation}
where $\ominus$ and $\oplus$ denote binary erosion and dilation, respectively.
These operations preserve the overall shape of signal occupancy while improving spatial connectivity, as illustrated in Fig.~\ref{fig: mo}.
{\name} applies one 2D closing, one 2D opening, and one horizontal 1D opening using fixed $3\times3$ and $1\times3$ all-one SEs.
Using small, fixed kernels ensures predictable runtime and avoids the exhaustive kernel search required by Searchlight~\cite{bell2023searchlight}.
In contrast to boundary inference with power rate change, binarization followed by MOs allows {\name} to focus on time--frequency proximity rather than absolute power variation for robust localization across diverse signal patterns.

%%%%%%%%%%%%%%%%%%%%%%%%%%%%%%%%%%%%%%%%%%%%%%%%%%%%%%%%%%%%%%%%%%%%%%%%%%%%%%%%
% MO example (move closer to text and first mention, also more balanced loc)
%%%%%%%%%%%%%%%%%%%%%%%%%%%%%%%%%%%%%%%%%%%%%%%%%%%%%%%%%%%%%%%%%%%%%%%%%%%%%%%%
%% figure begins
\begin{figure}[!t]
    \centering
    \includegraphics[width=0.98\columnwidth]{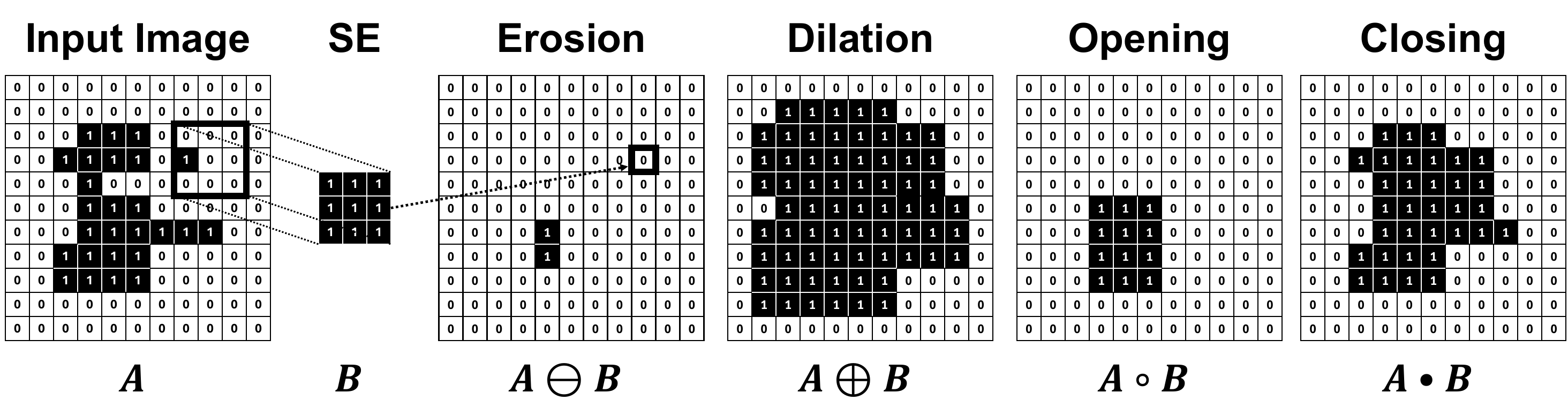}
    % \vspace{-1.5mm}
    \caption{Example MOs with a $3\times 3$ structuring element (SE).}
    \label{fig: mo}
    \vspace{-3mm}
\end{figure}
%% figure ends

%%%%%%%%%%%%%%%%%%%%%%%%%%%%%%%%%%%%%%%%%%%%%%%%%%%%%%%%%%%%%%%%%%%%%%%%%%%%%%%%
\myparatight{Bounding box generation.}
After morphological consolidation, each connected foreground region in the binary TF mask corresponds to a candidate signal occupancy.
{\name} applies standard connected component labeling (CCL) to efficiently identify individual energy blocks and extract their bounding boxes.
Specifically, the binarized image is scanned once, and connected foreground pixels are grouped into components using four-neighbor connectivity.
For each component, the minimum and maximum indices along the time and frequency axes define the bounding box.
Visited pixels are marked to avoid repeated labeling.
CCL yields one bounding box per detected energy block, as illustrated in Fig.~\ref{fig: sys_flow}(f).
Note that this stage can also be extended to output the boundaries of energy blocks when signal shapes are of interest.

%%%%%%%%%%%%%%%%%%%%%%%%%%%%%%%%%%%%%%%%%%%%%%%%%%%%%%%%%%%%%%%%%%%%%%%%%%%%%%%%
%%%%%%%%%%%%%%%%%%%%%%%%%%%%%%%%%%%%%%%%%%%%%%%%%%%%%%%%%%%%%%%%%%%%%%%%%%%%%%%%
\subsection{Algorithmic Time Complexity Analysis}
\label{subsec: time_complexity}

We analyze the end-to-end time complexity of {\name} with respect to the TF plot dimensions $(T, F)$.
Transforming streaming I/Q samples into a TF representation requires $T$ FFTs of size $F$, resulting in a complexity of $\bigO(TF\log F)$, which dominates the pipeline.
Subsequent image-based processing steps--including PSD aggregation, adaptive Otsu binarization, morphological operations with fixed kernels, and connected component labeling—each traverse the TF plot once and incur linear complexity $\bigO(TF)$.
Overall, {\name} achieves an end-to-end time complexity of $\bigO(TF\log F)$, enabling scalable wideband sensing under real-time constraints.

In contrast, Searchlight~\cite{bell2023searchlight} performs an exhaustive search over convolution kernels of varying time--frequency dimensions to localize arbitrarily sized energy blocks.
Without prior knowledge of signal dimensions, this matched-filter-style search leads to a worst-case complexity of $\bigO(T^2F^2)$.
Although the search space can be truncated heuristically, doing so either sacrifices detection generality or assumes prior signal knowledge, both of which limit real-time applicability.

While favorable asymptotic complexity is necessary for scalability, it is not sufficient to guarantee real-time operation.
Practical real-time performance further depends on system-level factors such as streaming architecture, task scheduling, and parallel execution, which we address in Sec.~\ref{subsec: sys_arch}.

%%%%%%%%%%%%%%%%%%%%%%%%%%%%%%%%%%%%%%%%%%%%%%%%%%%%%%%%%%%%%%%%%%%%%%%%%%%%%%%%
%%%%%%%%%%%%%%%%%%%%%%%%%%%%%%%%%%%%%%%%%%%%%%%%%%%%%%%%%%%%%%%%%%%%%%%%%%%%%%%%
\subsection{Real-time Architecture}
\label{subsec: sys_arch}

%% figure begins
\begin{figure}[!t]
    \centering
    \includegraphics[width=0.98\columnwidth]{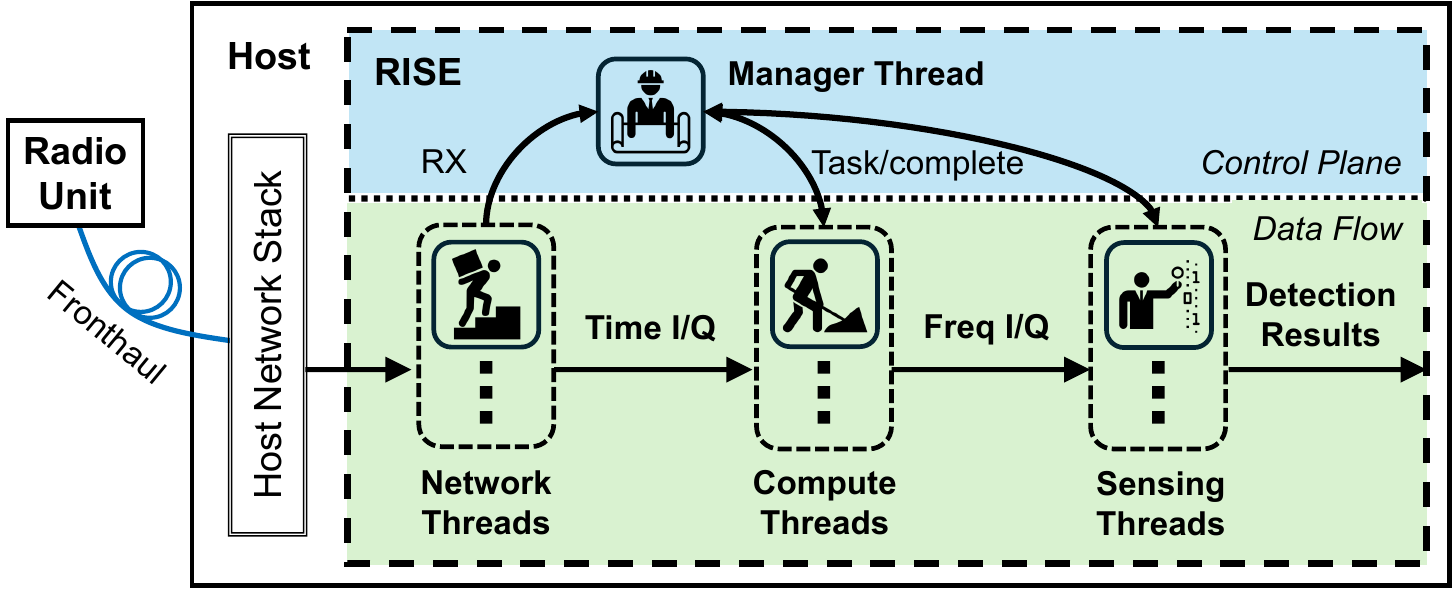}
    % \vspace{-1.5mm}
    \caption{Manager--worker architecture of {\name}, a multi-threaded streaming and processing system for real-time spectrum sensing.}
    \label{fig: sys_arch}
    \vspace{-3mm}
\end{figure}
%% figure ends

While {\name}'s image-processing pipeline admits favorable asymptotic complexity, real-time spectrum sensing imposes stricter constraints than algorithmic scalability alone.
Continuous wideband streaming and bounded-latency requirements restrict how computation can be parallelized and scheduled.
This section presents the system architecture of {\name}, which translates the algorithmic structure into a streaming, multi-threaded execution model for real-time operation.

Following recent work on vRAN PHY processing~\cite{ding2020agora, qi2024savannah}, {\name} is designed under strict streaming and bounded-latency constraints without specialized hardware, where algorithm structure directly dictates feasible parallelization strategies.
Fig.~\ref{fig: sys_arch} depicts the resulting system architecture.
{\name} can be flexibly deployed on a single multi-core general-purpose computer, with fronthaul I/Q streams from an external radio unit.
To decouple data arrival, spectral transformation, and 2D image processing, {\name} adopts a manager–worker model with four thread types: a manager thread, network threads, compute threads, and sensing threads.
Network threads stream I/Q samples from the radio, compute threads perform FFT and buffer frequency-domain results, and sensing threads execute the pipeline in Sec.~\ref{subsec: imag_proc} on TF plots.
The manager thread coordinates worker execution via concurrent queues.

\myparatight{Task scheduling.}
{\name} maintains a set of concurrent queues for the control message passing between the manager and worker threads.
The system starts as the network threads receive the input I/Q samples from the radio unit and notify the manager via the RX complete queue.
Compute threads and sensing threads each maintain a dedicated pair of task and completion queues, which are shared among all threads of the same type.
Each worker thread continuously polls its task queue, executes the assigned task, and reports completion to the manager via the corresponding completion queue.
Based on completed tasks, the manager thread schedules subsequent tasks by enqueuing them into the appropriate task queues.

\myparatight{Data arrangement and streaming interface.}
The intermediate data between each stage are stored in a global shared buffer.
The time-domain I/Q samples received by the network threads are first put into the buffer, and their time--frequency information is labeled for further processing.
{\name} performs $F$-point FFT on each chunk of the time-domain I/Q samples in the compute threads upon reception.
Each FFT result contributes a horizontal line to the TF plot, and the FFT size determines the plot width.
A ping-pong buffer serves as a synchronization boundary between time-parallel FFT execution and 2D-coupled image processing.
The buffer depth $T$ simultaneously determines the observation window and the end-to-end latency constraint, making it a first-class design parameter.
The TF plot is then fed to the sensing threads for image processing-based energy detection and localization.

%%%%%%%%%%%%%%%%%%%%%%%%%%%%%%%%%%%%%%%%%%%%%%%%%%%%%%%%%%%%%%%%%%%%%%%%%%%%%%%%
\myparatight{Real-time processing requirements.}
While FFT dominates the asymptotic complexity in Sec.~\ref{subsec: time_complexity}, real-time feasibility in {\name} is governed by whether TF-level processing can complete within a fixed window deadline.
As mentioned in Sec.~\ref{subsec: rt_challenge}, a real-time sensing system must process incoming data at a rate no slower than the input stream to avoid backlog accumulation, and {\name} targets \emph{bounded-latency processing} while sustaining the frontend input throughput.
At runtime, {\name} carries out three major stages: \textit{(i)} streaming raw I/Q samples from the radio, \textit{(ii)} transforming them into a TF plot in dimension $(T, F)$ via FFT, and \textit{(iii)} performing image-based energy detection and localization.
To avoid input loss, FFT processing must keep pace with the arrival of I/Q samples.
Given an FFT size of $F$ and a time resolution of $\Delta t = F / f_s$, each FFT must complete within a latency $\lat_{\text{FFT}}  = \Delta t$, and the corresponding throughput $\tput_{\text{FFT}} = F\cdot\text{size}_b(\tau)/{\Delta t}$, where $\text{size}_b(\tau)$ denotes the bit width of the data type $\tau$ used in processing.
Under the configuration used in this work ($F=1{,}024$, $\Delta t={10.24}\thinspace{\text{\usec}}$), FFT throughput requirement $\tput_{\text{FFT}}$ is {6.4}\thinspace{Gbps} when using 32-bit floating-point I/Q samples, exceeding the {3.2}\thinspace{Gbps} fronthaul input rate with 16-bit fixed-point samples.
Energy detection operates on TF plots constructed by accumulating FFT results in a ping-pong buffer of height $T$.
This buffering decouples FFT computation from image-based processing while defining the latency budget for TF plot-level computation.
Specifically, the latency ($\lat_{TF}$) and throughput ($\tput_{TF}$) requirements for processing each TF plot (Fig.~\ref{fig: sys_flow}) are
\begin{equation}
\textstyle % force sigma on the right.
\lat_{TF} = T \cdot \Delta t
,\quad
\tput_{TF}
= \frac{TF\cdot \text{size}_b(\tau)}{T\cdot \Delta t}
= \frac{F\cdot\text{size}_b(\tau)}{\Delta t}.
\end{equation}
Processing a TF plot must be completed before the next plot becomes available for uninterrupted sensing.
Note that $\lat_{TF}$ is proportional to the TF plot height $T$ and the time resolution $\Delta t$, while $\tput_{TF}$ is inversely proportional to $\Delta t$ but independent of $T$.
Although $\tput_{TF}$ equals the FFT-stage throughput ($\tput_{\mathrm{TF}} = \tput_{\text{FFT}}$), the latency budgets differ ($\lat_{\mathrm{TF}}\neq\lat_{\text{FFT}}$): FFT operates at the granularity of individual time segments ($\lat_{\text{FFT}} = {10.24}\thinspace{\text{\usec}}$), whereas TF-level processing is constrained by the accumulation window ($\lat_{\text{FFT}} = {20.48}\thinspace{\text{\msec}}$).
Here, the TF plot height $T$ is a system-level design choice driven by application requirements (e.g., temporal resolution and sensing delay), rather than an algorithmic hyperparameter.
{\name} is designed to satisfy bounded-latency processing for any configured $T$, provided sufficient parallelism is available.

%%%%%%%%%%%%%%%%%%%%%%%%%%%%%%%%%%%%%%%%%%%%%%%%%%%%%%%%%%%%%%%%%%%%%%%%%%%%%%%%
\myparatight{Task partitioning for parallelism.}
Although {\name} is implemented as a software system on a multi-core server, achieving real-time performance is not a matter of naive thread parallelization.
Several stages in the pipeline exhibit inherent data dependencies across time and frequency, which fundamentally constrain task partitioning.
Specifically, FFT processing is parallel across time segments and is therefore time-parallelized across compute threads.
In contrast to FFT processing, which is naturally \emph{time-parallel} across incoming I/Q chunks, TF-domain processing exhibits a different parallel structure.
PSD estimation and adaptive thresholding aggregate samples over time but are independent across frequency components, making them amenable to \emph{frequency-parallel} execution.
This necessitates a transition from time-parallel FFT processing to frequency-oriented TF processing via buffered TF plots.
Subsequent stages, including MOs and CCL, operate on two-dimensional spatial adjacency but can still be partitioned along the frequency axis with appropriate boundary handling.

To bridge the time-streaming FFT stage and the TF-domain processing stage, {\name} employs a ping-pong buffer as a domain-crossing boundary.
This design decouples FFT throughput from image-processing latency, enabling concurrent writes by compute threads and reads by sensing threads while enforcing a bounded processing window.
Under ideal parallelism across both time-parallel FFT processing and frequency-parallel TF-domain stages, the resulting hybrid partitioning reduces the effective critical path to $\bigO(F\log F + T)$.
More importantly, this design decouples algorithmic complexity from real-time guarantees, enabling predictable bounded-latency execution under sustained input throughput.

\section{Experiment Setup and Baselines}
\label{sec: impl}

This section details the experiment environment to evaluate {\name} and the implementation of baseline algorithms.

%%%%%%%%%%%%%%%%%%%%%%%%%%%%%%%%%%%%%%%%%%%%%%%%%%%%%%%%%%%%%%%%%%%%%%%%%%%%%%%%
%%%%%%%%%%%%%%%%%%%%%%%%%%%%%%%%%%%%%%%%%%%%%%%%%%%%%%%%%%%%%%%%%%%%%%%%%%%%%%%%
\subsection{{\name} Experiment Setup}\label{subsec: exp_setup}

%%%%%%%%%%%%%%%%%%%%%%%%%%%%%%%%%%%%%%%%%%%%%%%%%%%%%%%%%%%%%%%%%%%%%%%%%%%%%%%%
\myparatight{Hardware/software platform.}
We implement the core functions of {\name} in about 2,700 lines of C++ code for real-time processing, and develop a suite of Python helper scripts for offline analysis and evaluation.
The multi-threaded architecture is implemented based on Agora~\cite{ding2020agora, qi2024savannah}, an open-source wireless PHY processing framework.
Arithmetic operations leverage the Armadillo library~\cite{sanderson2016armadillo} to exploit instruction-level parallelism on modern CPUs.
We deploy {\name} on a Dell PowerEdge R750 server with a 56-core Intel Xeon Gold 6348 CPU @{2.6}\thinspace{GHz} and Ubuntu {20.04.6} LTS.
Detection outputs are stored in \texttt{.csv} format for each subband and time interval.

%%%%%%%%%%%%%%%%%%%%%%%%%%%%%%%%%%%%%%%%%%%%%%%%%%%%%%%%%%%%%%%%%%%%%%%%%%%%%%%%
\myparatight{Signal generation.}
We use Sig-Gen, an open-source MATLAB-based signal generator from RFSynth~\cite{sankar2024rfsynth}, to produce synthetic I/Q samples with ground-truth annotations.
Sig-Gen takes a single \texttt{\small .yaml} configuration file to define multiple signals within one spectrum snapshot, where each signal is specified with its own signal type (e.g., OFDM/DSSS Wi-Fi, BLE, or thermal Gaussian white noise), center frequency, transmitter power in dBm, and temporal occurrence.
Receiver parameters, such as the sampling rate $f_s$ and RX center frequency, are also specified in the same configuration file.
Sig-Gen incorporates hardware impairments (e.g., I/Q mismatch, frequency offset, and DC offset) and wireless channel effects, yielding high-fidelity waveforms.
It outputs raw I/Q samples in \texttt{\small .32cf} format along with metadata in a \texttt{\small .json} file, which records the start and end times and frequencies of each energy block for all configured signals.
We use the raw I/Q samples as input to {\name} and the metadata as ground truth for evaluation.
Unless otherwise noted, all experiments target a {100}\thinspace{MHz} wideband sensing, which defines the TF plot dimensions and real-time processing requirements throughout the paper.

%%%%%%%%%%%%%%%%%%%%%%%%%%%%%%%%%%%%%%%%%%%%%%%%%%%%%%%%%%%%%%%%%%%%%%%%%%%%%%%%
%% figure begins
\begin{figure}[!t]
    \centering
    \includegraphics[width=0.98\columnwidth]{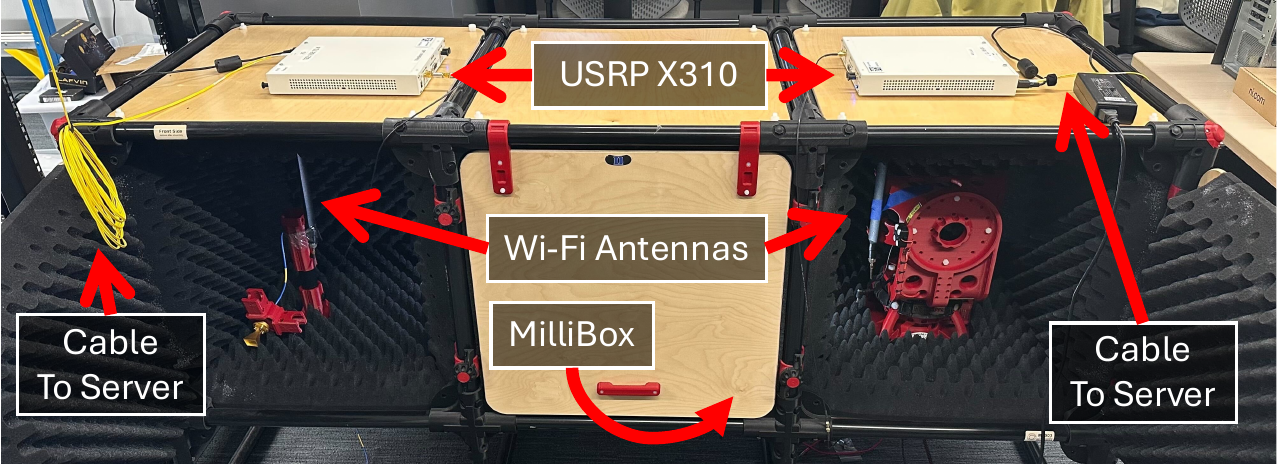}
    \caption{OTA setup for an interference-controlled environment.}
    \label{fig: ota_setup}
    \vspace{-3mm}
\end{figure}
%% figure ends

%%%%%%%%%%%%%%%%%%%%%%%%%%%%%%%%%%%%%%%%%%%%%%%%%%%%%%%%%%%%%%%%%%%%%%%%%%%%%%%%
% Grand evaluation figure (brought forward)
%%%%%%%%%%%%%%%%%%%%%%%%%%%%%%%%%%%%%%%%%%%%%%%%%%%%%%%%%%%%%%%%%%%%%%%%%%%%%%%%
%% figure begins
\begin{figure*}[!t]
    \centering
    \vspace{-3mm}
    \begin{minipage}[t]{0.9\textwidth}
    \centering
    % row 1
    \subfloat[Sparse (IoU 63.38\%)]{
    \includegraphics[width=0.27\columnwidth]{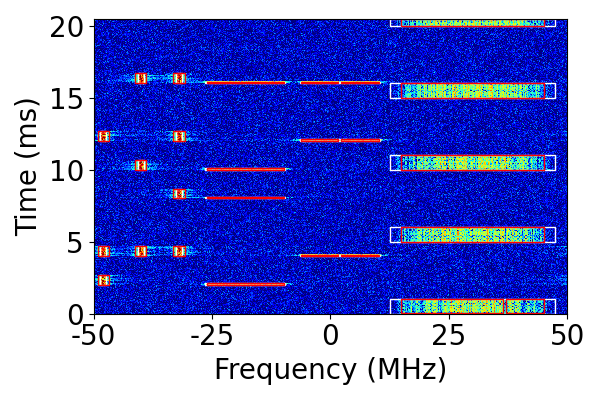}
    \label{subfig: sparse}}
    \hspace{1mm}
    \subfloat[RFSynth Default (IoU 72.23\%)]{
    \includegraphics[width=0.27\columnwidth]{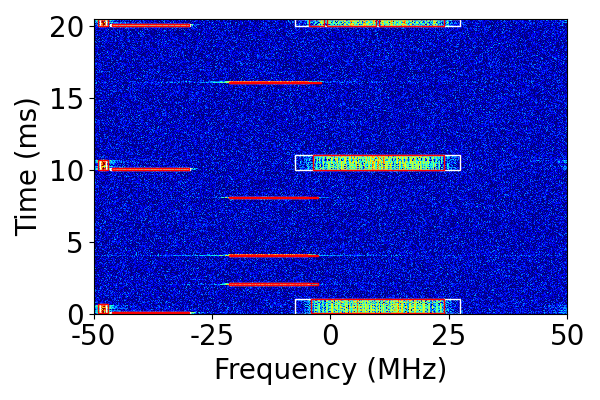}
    \label{subfig: test}}
    \hspace{1mm}
    \subfloat[Control (IoU 48.42\%)]{
    \includegraphics[width=0.27\columnwidth]{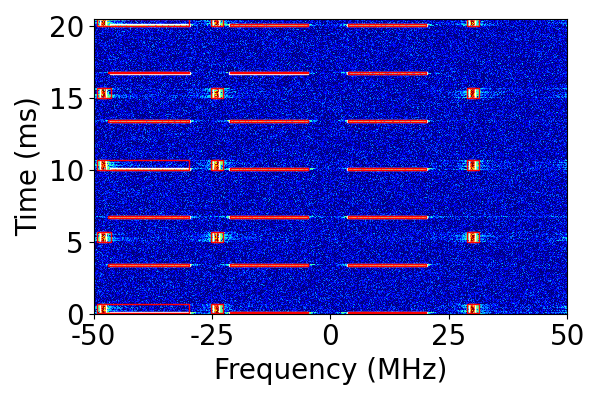}
    \label{subfig: std}}

    % row 2
    \vspace{-3mm}
    \subfloat[Dense, Mixed Signals (IoU 60.46\%)]{
    \includegraphics[width=0.27\columnwidth]{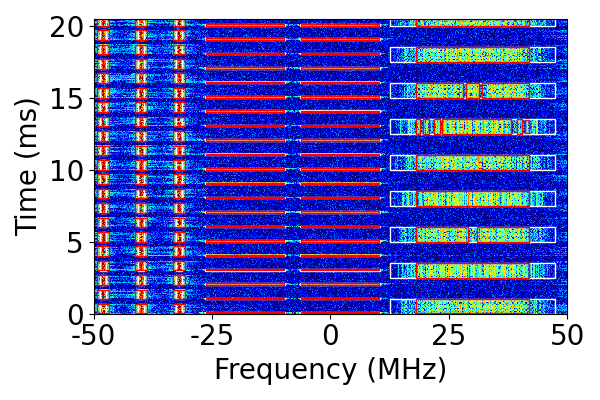}
    \label{subfig: dense}}
    \hspace{1mm}
    \subfloat[Dense, DSSS Only (IoU 66.38\%)]{
    \includegraphics[width=0.27\columnwidth]{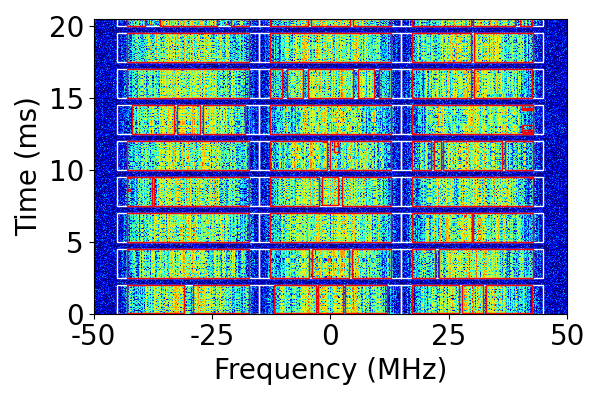}
    \label{subfig: dense_ds3}}
    \hspace{1mm}
    \subfloat[Wideband ({500}\thinspace{MHz}, IoU 66.18\%)]{
    \includegraphics[width=0.27\columnwidth]{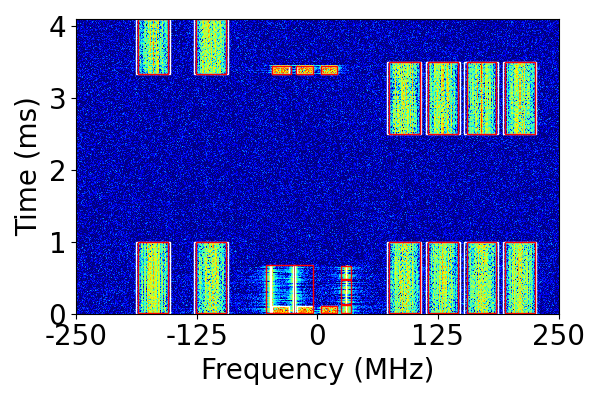}
    \label{subfig: wb}}
    \end{minipage}
    \hspace{-10mm}
    % colorbar
    \raisebox{-53mm}{
    \includegraphics[height=0.5\columnwidth]{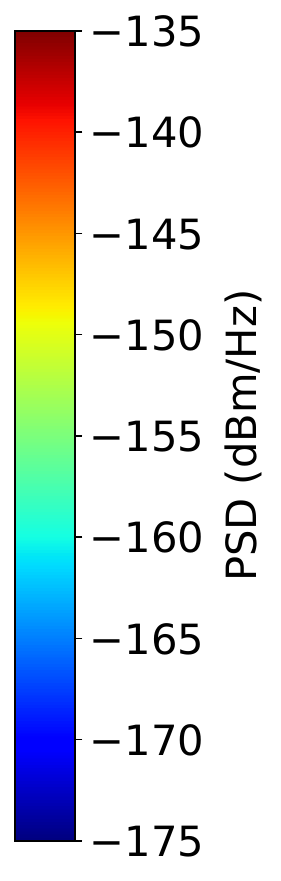}
    }
    \caption{
    Representative scenarios used in both simulation and OTA evaluations.
    Simulation-based results for signal detection and localization are shown.
    Ground-truth bounding boxes are shown in white, and the detection results are shown in red.}
    \label{fig: result_multi-config}
    \vspace{-3mm}
\end{figure*}
%% figure ends

%%%%%%%%%%%%%%%%%%%%%%%%%%%%%%%%%%%%%%%%%%%%%%%%%%%%%%%%%%%%%%%%%%%%%%%%%%%%%%%%
\myparatight{Over-the-air (OTA) setup and effective SNR.}
We evaluate {\name} using OTA measurements collected with USRP X310 SDRs (configured with {15}\thinspace{dB} RX gain, $f_s={100}\thinspace{\mathrm{MSps}}$, and centered at {3}\thinspace{GHz}) with commercial Wi-Fi antennas, in controlled experiments inside an RF-shielded enclosure (millibox) shown in Fig.~\ref{fig: ota_setup}.
To derive ground-truth TF labels from received traces, we embed samples from RFSynth into structured waveforms with known temporal offsets.
Relative signal strengths of coexisting technologies are normalized to reflect representative power disparities, while the absolute SNR is controlled via TX gain and fixed RF attenuators.
The effective SNR is computed as the ratio of signal power within ground-truth-labeled TF regions to noise power measured during silence.
This setup enables controlled SNR sweeps under practical OTA conditions and quantitative comparison against ground truth.

%%%%%%%%%%%%%%%%%%%%%%%%%%%%%%%%%%%%%%%%%%%%%%%%%%%%%%%%%%%%%%%%%%%%%%%%%%%%%%%%
%%%%%%%%%%%%%%%%%%%%%%%%%%%%%%%%%%%%%%%%%%%%%%%%%%%%%%%%%%%%%%%%%%%%%%%%%%%%%%%%
\subsection{Baseline Implementation}\label{subsec: baseline_impl}
We compare {\name} against representative spectrum sensing baselines, including Searchlight~\cite{bell2023searchlight} as the energy-detection-based method, and DeepRadar~\cite{sarkar2021deepradar} and U-Net~\cite{uvaydov2024stitching} as learning-based object detection and image segmentation approaches.

%%%%%%%%%%%%%%%%%%%%%%%%%%%%%%%%%%%%%%%%%%%%%%%%%%%%%%%%%%%%%%%%%%%%%%%%%%%%%%%%
\myparatight{Searchlight (convolution-based energy detection).}
We implement the core Searchlight~\cite{bell2023searchlight} detection pipeline (Sec.~\ref{subsec: searchlight_flow}) in Python to enable algorithm-level comparison independent of system-level optimizations.
For fair comparison, Searchlight operates on the same TF plot representation as {\name}.
To mitigate the otherwise prohibitive cost of exhaustive kernel search, we restrict the convolution kernel height and width to powers of two, reducing the worst-case complexity from $\bigO(T^{2}F^{2})$ to $\bigO(TF\log T\log F)$.

%%%%%%%%%%%%%%%%%%%%%%%%%%%%%%%%%%%%%%%%%%%%%%%%%%%%%%%%%%%%%%%%%%%%%%%%%%%%%%%%
\myparatight{DeepRadar (ML-based object detection).}
DeepRadar~\cite{sarkar2021deepradar} is a YOLO-based object detection framework designed for CBRS radar signal detection.
We implement DeepRadar in PyTorch and retrain it using TF plots synthesized by RFSynth~\cite{sankar2024rfsynth}, covering Wi-Fi and BLE signals over a {100}\thinspace{MHz} band.
We apply minimal and standard adaptations to the DeepRadar model to enable bounding box prediction on TF plots, including a convolutional prediction head, adaptive pooling for skewed TF dimensions, conventional box regression and confidence losses, and non-maximum suppression for post-processing.

%%%%%%%%%%%%%%%%%%%%%%%%%%%%%%%%%%%%%%%%%%%%%%%%%%%%%%%%%%%%%%%%%%%%%%%%%%%%%%%%
\myparatight{U-Net (ML-based image segmentation).}
Spectrum sensing can also be formulated as an image segmentation problem, and a U-Net with self-attention has been proposed to estimate spectrum occupancy at the pixel level, together with an open-source implementation and dataset pipeline~\cite{uvaydov2024stitching}.
The pretrained model does not generalize to our data due to strong positional and energy continuity priors in the original dataset (i.e., signals centered in frequency and occupying contiguous bands).
We therefore retrain the model using RFSynth-generated signals under the same configurations as the other ML baseline to ensure a fair comparison, with ground truth labels adapted from bounding-box annotations to pixel-level occupancy maps.
The model operates on per-FFT spectrum snapshots and outputs per-class signal occupancy maps.
Following the original model design~\cite{uvaydov2024stitching}, which operates on a {25}\thinspace{MHz} bandwidth, wideband signals are partitioned into independent subbands.
Both training/inference and ground-truth annotations follow the same partitioning, and detection performance is evaluated consistently across all subbands using pixel-wise accumulation into a single spectrum-occupancy map.
U-Net outputs pixel-wise binary segmentation masks that represent signal occupancy on the TF plot.
Although these predictions are not detection boxes, the resulting $P_d$ and IoU remain comparable to box-based metrics, as both quantify the overlap between predicted and ground-truth occupied regions.
In contrast, $P_{fa}$ in {\eqref{eq_pd_pfa}} is defined over detection boxes and is therefore not directly applicable to pixel-level predictions.

\section{Evaluation}
\label{sec: eval}

We evaluate {\name} for detection performance and real-time capability, and compare it against representative baselines including Searchlight~\cite{bell2023searchlight}, DeepRadar~\cite{sarkar2021deepradar}, and U-Net~\cite{uvaydov2024stitching}.

%%%%%%%%%%%%%%%%%%%%%%%%%%%%%%%%%%%%%%%%%%%%%%%%%%%%%%%%%%%%%%%%%%%%%%%%%%%%%%%%
\myparatight{Evaluation setup.}
Unless otherwise noted, all results are evaluated on TF plots generated from received I/Q samples at $f_s={100}\thinspace\mathrm{MSps}$ (corresponding to {100}\thinspace{MHz} sensing bandwidth) centered at {2.45}\thinspace{GHz}.
Detection and image processing operate on FFT-based spectral magnitudes (i.e., absolute values, referred to as PSD in Sec.~\ref{sec: design}) without explicit normalization; for visualization only, these values are calibrated by a noise floor estimate from low-power samples and expressed in dBm/Hz for a physically meaningful power scale.
With $N_{\text{FFT}}=1{,}024$, the TF plots have frequency and time resolutions of $\Delta f \approx 98\thinspace\mathrm{kHz}$ and $\Delta t = 10.24\thinspace\mu\mathrm{s}$; the dimension $(T,F)=(2000,1024)$ corresponds to a sensing window of {20.48}\thinspace{ms}.
The signals include BLE and Wi-Fi (OFDM/DSSS) with diverse time--frequency occupancy patterns.

%%%%%%%%%%%%%%%%%%%%%%%%%%%%%%%%%%%%%%%%%%%%%%%%%%%%%%%%%%%%%%%%%%%%%%%%%%%%%%%%
%%%%%%%%%%%%%%%%%%%%%%%%%%%%%%%%%%%%%%%%%%%%%%%%%%%%%%%%%%%%%%%%%%%%%%%%%%%%%%%%
\subsection{Detection Performance}
\label{subsec: det_perf}

We evaluate the detection performance of {\name} using representative scenarios under varying IoU thresholds and SNR.

%%%%%%%%%%%%%%%%%%%%%%%%%%%%%%%%%%%%%%%%%%%%%%%%%%%%%%%%%%%%%%%%%%%%%%%%%%%%%%%%
\subsubsection{Representative detection scenarios}
We evaluate the image processing pipeline of {\name} (Sec.~\ref{subsec: imag_proc}) on a set of representative scenarios using the detection metrics defined in Sec.~\ref{sub2sec: detect_metric}.
Fig.~\ref{fig: result_multi-config} presents six scenarios operating at fixed, nominal SNRs ({15}\thinspace{dB} for BLE and {20/14}\thinspace{dB} for OFDM/DSSS Wi-Fi) with fixed per-technology bandwidths ({2}\thinspace{MHz} for BLE and {20/30--35}\thinspace{MHz} for OFDM/DSSS Wi-Fi), but varying signal density, i.e., the temporal and spectral occupancy patterns.
In the figure, ground-truth and detected energy blocks are marked by white and red bounding boxes, respectively; a well-aligned detection is visually red-dominant.
The per-scenario IoU is computed by averaging, over all ground-truth boxes, the IoU with their best-matched (maximum IoU) detections.

\myparatight{Sparse scenarios.}
The top row of Fig.~\ref{fig: result_multi-config} evaluates temporally sparse coexistence.
In Fig.~\ref{fig: result_multi-config}\subref{subfig: sparse}, {\name} reliably detects BLE and OFDM Wi-Fi signals but occasionally fragments DSSS Wi-Fi into multiple detections.
This results in an IoU of 63.38\%, with $(P_d, P_{fa})=(63.63\%, 33.33\%)$ at $\thresIoU=0.5$.
The primary source of IoU degradation is boundary mismatch: {\name} estimates signal ranges based on energy above the noise floor, whereas the ground truth is defined at the {3}\thinspace{dB} points.
Consequently, narrowband BLE detections tend to be wider than ground truth, while DSSS detections are narrower, despite sharing similar centers.
Figs.~\ref{fig: result_multi-config}\subref{subfig: test}\subref{subfig: std} illustrate the default RFSynth configuration and a standard coexistence scenario with BLE operating on its control channel and Wi-Fi occupying non-overlapping channels.
Although closely spaced signals may merge into a single detection when spectral overlap occurs, all dominant energy peaks are successfully identified.

\myparatight{Dense scenarios.}
Figs.~\ref{fig: result_multi-config}\subref{subfig: dense}\subref{subfig: dense_ds3} evaluate dense coexistence, where {\name} distinguishes closely packed signals, at the cost of occasional over-segmentation for sparse DSSS waveforms.

\myparatight{Wideband scenario.}
Fig.~\ref{fig: result_multi-config}\subref{subfig: wb} further demonstrates {500}\thinspace{MHz} bandwidth sensing at $f_s=500$\thinspace{MSps}, resulting in finer time resolution of $\Delta t=2.048\thinspace\mu$s (with $N_{\text{FFT}}=1024$) and a sensing window $T\cdot \Delta t={4.096}\thinspace{\mathrm{ms}}$.
This scenario mirrors the configuration in Fig.~\ref{fig: result_multi-config}\subref{subfig: std} for the central {100}\thinspace{MHz} band, while additional DSSS signals occupy the outer spectrum.

%%%%%%%%%%%%%%%%%%%%%%%%%%%%%%%%%%%%%%%%%%%%%%%%%%%%%%%%%%%%%%%%%%%%%%%%%%%%%%%%

%% figure begins
\begin{figure}[!t]
    \centering
    % \vspace{-5mm}
    \subfloat{
    \includegraphics[width=0.8\columnwidth]{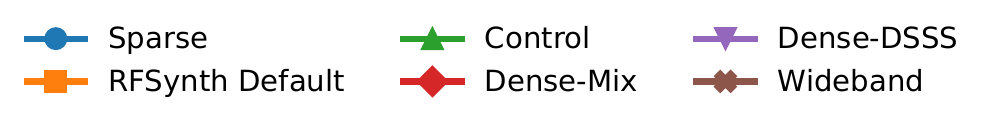}}
    \vspace{-4mm}
    \subfloat{
    \includegraphics[width=0.45\columnwidth]{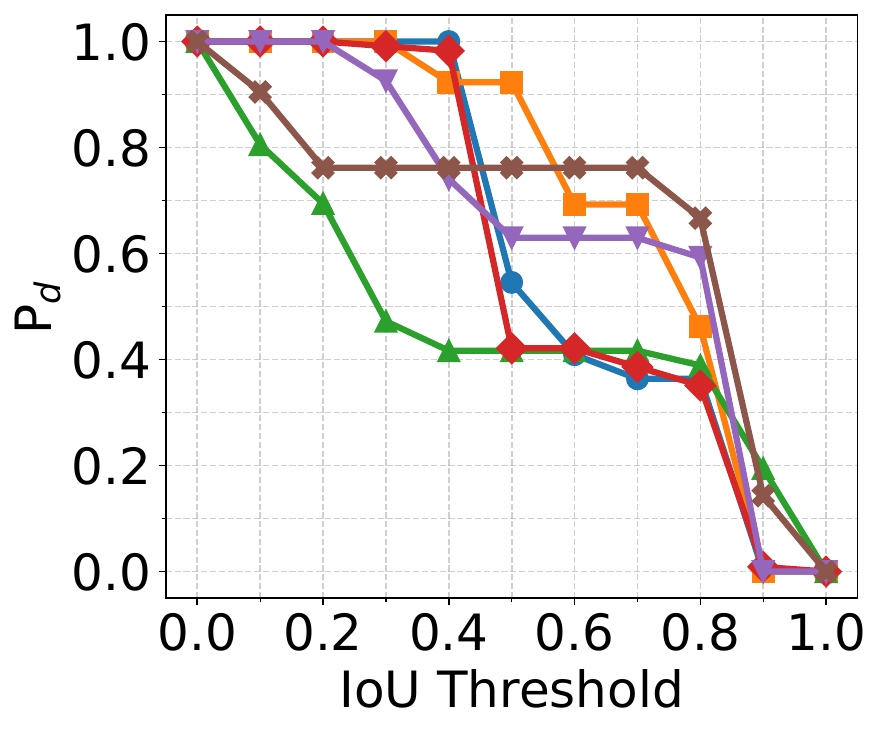}
    \label{subfig: tiou_pd}}
    \subfloat{
    \includegraphics[width=0.45\columnwidth]{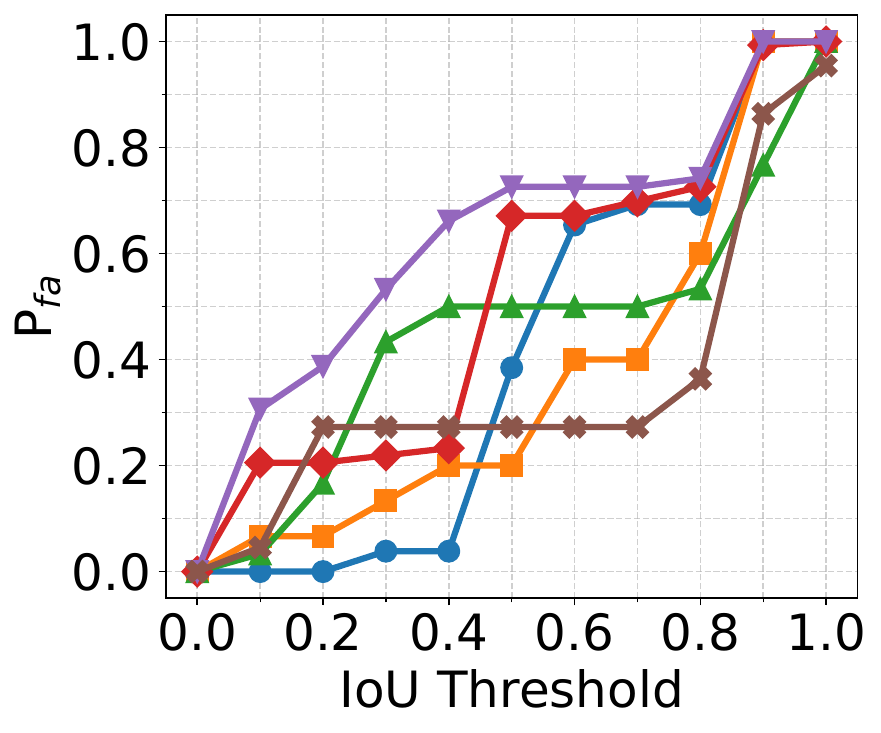}
    \label{subfig: tiou_pfa}}
    \vspace{-1.5mm}
    \caption{Probability of detection $P_d$ and false alarm rate $P_{fa}$ as functions of IoU threshold $\thresIoU$ for representative scenarios.}
    \label{fig: tiou_pd_pfa}
    \vspace{-3mm}
\end{figure}
%% figure ends

%%%%%%%%%%%%%%%%%%%%%%%%%%%%%%%%%%%%%%%%%%%%%%%%%%%%%%%%%%%%%%%%%%%%%%%%%%%%%%%%
% Simulation SNR (brought forward)
%%%%%%%%%%%%%%%%%%%%%%%%%%%%%%%%%%%%%%%%%%%%%%%%%%%%%%%%%%%%%%%%%%%%%%%%%%%%%%%%
%% figure begins
\begin{figure}[!t]
    \centering
    % \vspace{-3mm}
    %  \subfloat{
    \includegraphics[width=0.48\columnwidth]{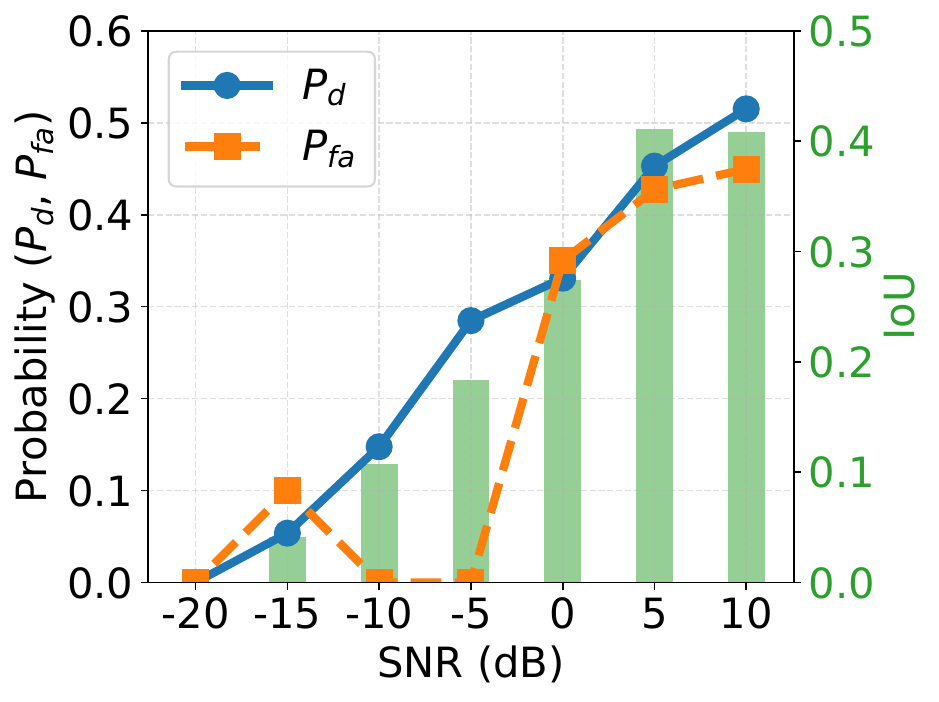}
    %  \label{subfig: snr_pd_pfa_iou_sim}}
    \vspace{-1.5mm}
    \caption{Average $P_d$, $P_{fa}$, and IoU vs. SNR in simulation ($\thresIoU=0.5$).}
    \label{fig: snr_pd_pfa_iou_sim}
    \vspace{-3mm}
\end{figure}
%% figure ends
%%%%%%%%%%%%%%%%%%%%%%%%%%%%%%%%%%%%%%%%%%%%%%%%%%%%%%%%%%%%%%%%%%%%%%%%%%%%%%%%
%%%%%%%%%%%%%%%%%%%%%%%%%%%%%%%%%%%%%%%%%%%%%%%%%%%%%%%%%%%%%%%%%%%%%%%%%%%%%%%%

%%%%%%%%%%%%%%%%%%%%%%%%%%%%%%%%%%%%%%%%%%%%%%%%%%%%%%%%%%%%%%%%%%%%%%%%%%%%%%%%
\subsubsection{Detection performance versus IoU threshold}
We next examine how $P_d$ and $P_{fa}$ vary as functions of $\thresIoU$, which is applied to the pairwise IoU between detected and ground-truth boxes in \eqref{eq_td_fd_by_thres_iou}.
As $P_d$ and $P_{fa}$ depend on the IoU requirement imposed by downstream applications, varying $\thresIoU$ exposes different operating points.
Fig.~\ref{fig: tiou_pd_pfa} plots $P_d$ and $P_{fa}$ versus $\thresIoU$ for each representative scenario in Fig.~\ref{fig: result_multi-config}.
The resulting trends differ across scenarios, reflecting their distinct time--frequency occupancy patterns.
In addition, relaxing the IoU threshold from $\thresIoU=0.5$ to $\thresIoU=0.4$ improves average $(P_d,  P_{fa})$ from (61.63\%, 45.91\%) to (80.42\%, 31.76\%), corresponding to an 18.79\% gain in $P_d$.
Note that the average IoU remains at $62.84\%$ across IoU thresholds, indicating that the observed trade-offs are primarily driven by the threshold criterion rather than changes in localization accuracy.

%%%%%%%%%%%%%%%%%%%%%%%%%%%%%%%%%%%%%%%%%%%%%%%%%%%%%%%%%%%%%%%%%%%%%%%%%%%%%%%%
% Real-time latency figure (brought forward)
%%%%%%%%%%%%%%%%%%%%%%%%%%%%%%%%%%%%%%%%%%%%%%%%%%%%%%%%%%%%%%%%%%%%%%%%%%%%%%%%
%% figure begins
\begin{figure}[!t]
    \centering
    \vspace{-1.5mm}
    % \subfloat{
    \includegraphics[width=0.96\columnwidth]{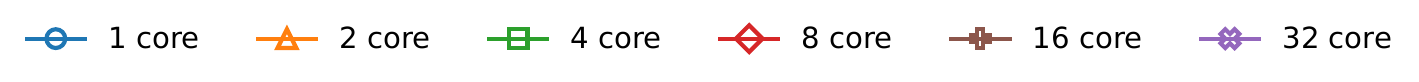}\vspace{-4mm}
    \subfloat[Single-threaded Time Breakdown]{
    \includegraphics[width=0.48\columnwidth]{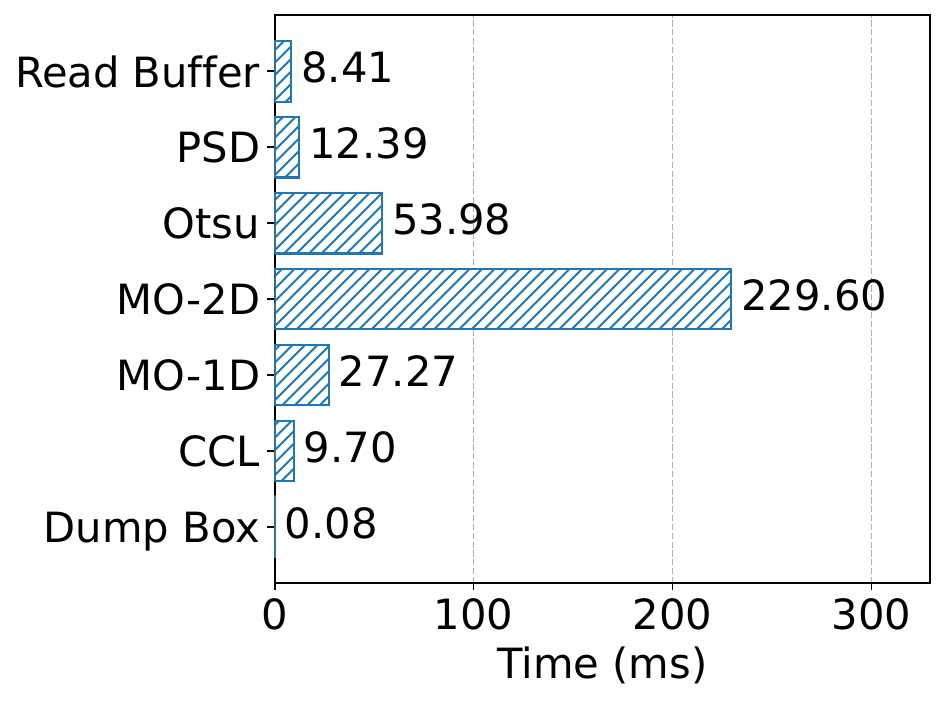}
    \label{subfig: time_breakdown}}
    \subfloat[Multi-threaded Time CCDF]{
    \includegraphics[width=0.48\columnwidth]{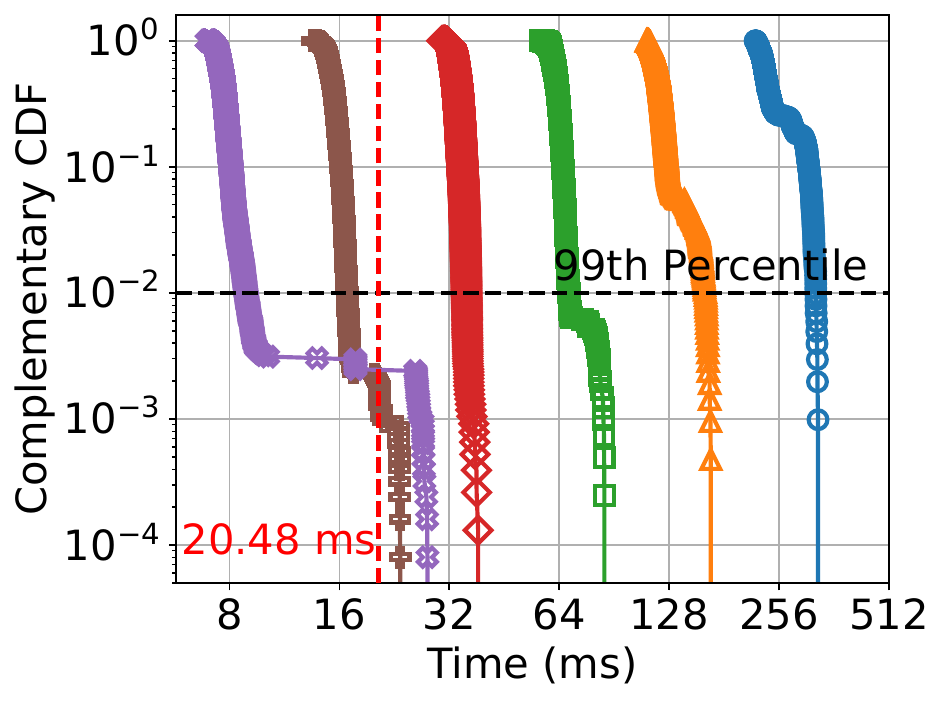}
    \label{subfig: ccdf_time_vs_subband}}
    % \subfloat[Band Partition]{
    % \includegraphics[width=0.48\columnwidth]{figs/time_vs_subband.pdf}
    % \label{subfig: time_vs_subband}}
    % \vspace{-1.5mm}
    \caption{The execution time of {\name} in C++ implementation. (a) The average time contributed by each step. (b) Parallelization enabled by multi-threading meets the deadline.}
    \label{fig: eval_time}
    \vspace{-1.5mm}
\end{figure}
%% figure ends
%%%%%%%%%%%%%%%%%%%%%%%%%%%%%%%%%%%%%%%%%%%%%%%%%%%%%%%%%%%%%%%%%%%%%%%%%%%%%%%%
%%%%%%%%%%%%%%%%%%%%%%%%%%%%%%%%%%%%%%%%%%%%%%%%%%%%%%%%%%%%%%%%%%%%%%%%%%%%%%%%

%%%%%%%%%%%%%%%%%%%%%%%%%%%%%%%%%%%%%%%%%%%%%%%%%%%%%%%%%%%%%%%%%%%%%%%%%%%%%%%%
% OTA lines (brought forward)
%%%%%%%%%%%%%%%%%%%%%%%%%%%%%%%%%%%%%%%%%%%%%%%%%%%%%%%%%%%%%%%%%%%%%%%%%%%%%%%%
%% figure begins
\begin{figure}[!t]
    \vspace{-3mm}
    \centering
    \includegraphics[width=0.8\columnwidth]{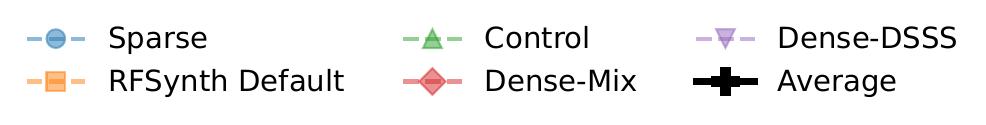}\vspace{-4mm}
    \subfloat[$P_d$ vs. SNR for each scenario]{
    \includegraphics[width=0.47\columnwidth]{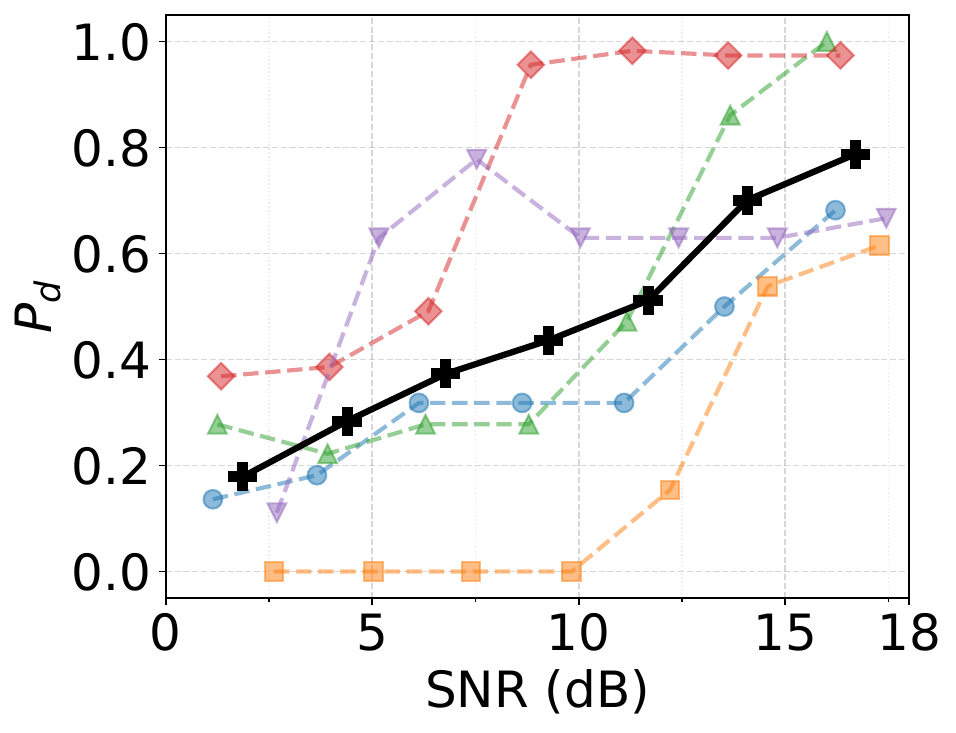}
    % figs/imag_proc_tf_box_ota-100mhz.png
    \label{subfig: snr_pd_ota}}
    \subfloat[Average $P_d$/$P_{fa}$/IoU vs. SNR]{
    \includegraphics[width=0.495\columnwidth]{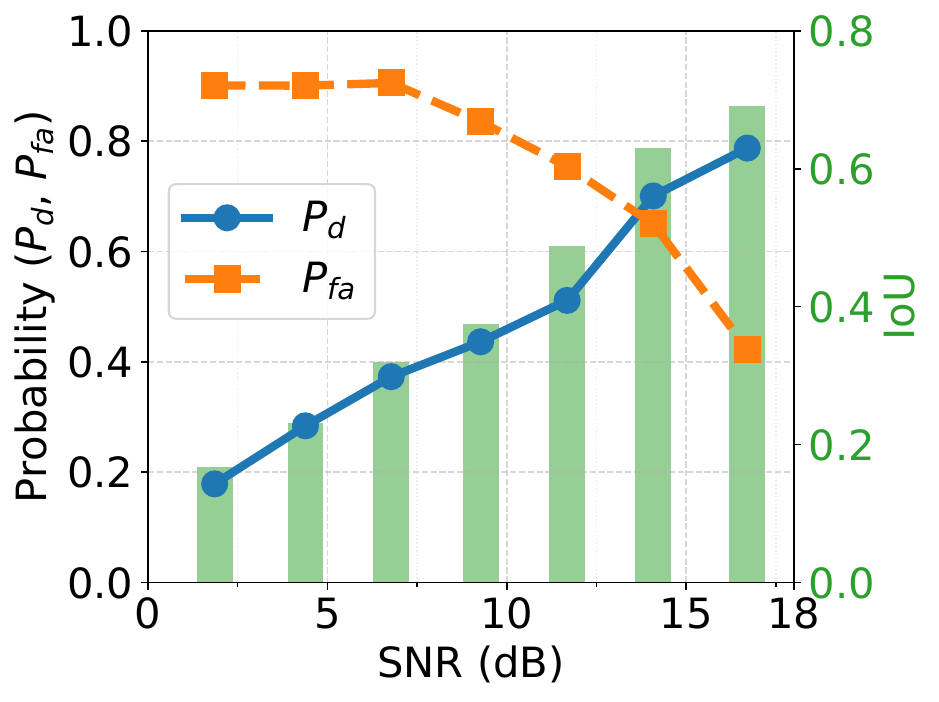}
    \label{subfig: snr_pd_pfa_iou_ota}}
    % \vspace{-1.5mm}
    % \caption{SNR sweep ($\thresIoU=0.5$) for OTA evaluaiton.}
    \caption{SNR sweep under controlled OTA evaluation ($\thresIoU=0.5$).}
    \label{fig: ota_controlled}
    \vspace{-3.5mm}
\end{figure}
%% figure ends
%%%%%%%%%%%%%%%%%%%%%%%%%%%%%%%%%%%%%%%%%%%%%%%%%%%%%%%%%%%%%%%%%%%%%%%%%%%%%%%%
%%%%%%%%%%%%%%%%%%%%%%%%%%%%%%%%%%%%%%%%%%%%%%%%%%%%%%%%%%%%%%%%%%%%%%%%%%%%%%%%

%%%%%%%%%%%%%%%%%%%%%%%%%%%%%%%%%%%%%%%%%%%%%%%%%%%%%%%%%%%%%%%%%%%%%%%%%%%%%%%%
% Baseline comparison bar chart (brought forward)
%%%%%%%%%%%%%%%%%%%%%%%%%%%%%%%%%%%%%%%%%%%%%%%%%%%%%%%%%%%%%%%%%%%%%%%%%%%%%%%%
%% figure begins
\begin{figure*}[!t]
    \centering
    % \vspace{-5mm}
    % \subfloat[]{
    % \includegraphics[width=0.44\columnwidth]{figs/imag_proc_tf_box_ota_yellow-red.png}
    % % figs/imag_proc_tf_box_ota-100mhz.png
    % \label{subfig: ota_itw}}
    % \subfloat[]{
    \includegraphics[width=1.9\columnwidth]{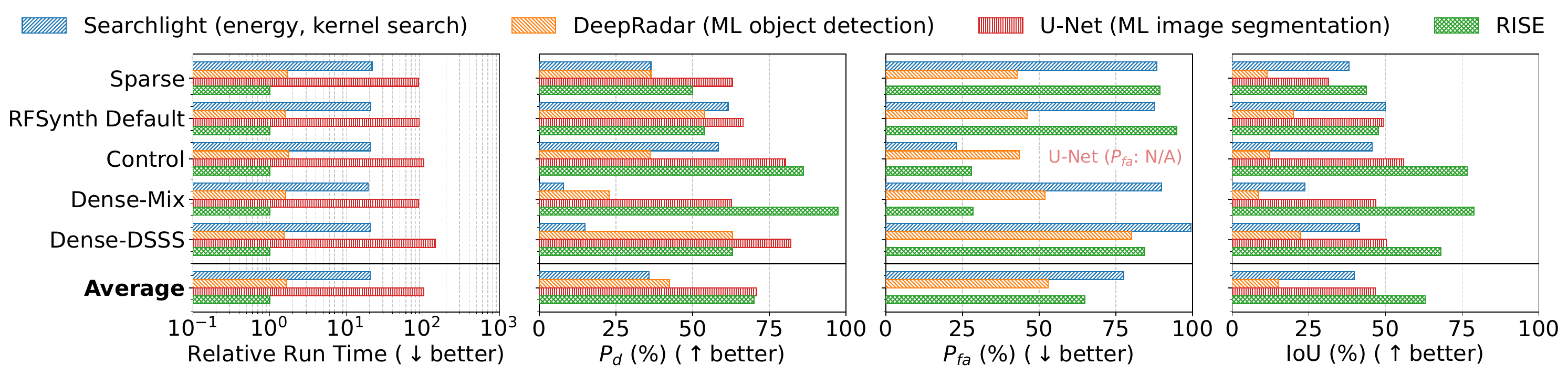}
    % \label{subfig: baseline_ota_time_pd_pfa_iou}}
    \vspace{-1.5mm}
    \caption{ %(a) OTA in-the-wild snapshot. (b) 
    Baseline comparison using OTA captures at an average effective SNR of {14.08}\thinspace{dB}.
    Searchlight~\cite{bell2023searchlight} and {\name} use $\thresIoU=0.5$, while DeepRadar~\cite{sarkar2021deepradar} uses $\thresIoU=0.2$ due to its sensitivity to stricter IoU thresholds.
    U-Net~\cite{uvaydov2024stitching} reports pixel-level $P_d$ and IoU; $P_{fa}$ not applicable.}
    \label{fig: baseline_ota}
    \vspace{-4.5mm}
\end{figure*}
%% figure ends
%%%%%%%%%%%%%%%%%%%%%%%%%%%%%%%%%%%%%%%%%%%%%%%%%%%%%%%%%%%%%%%%%%%%%%%%%%%%%%%%
%%%%%%%%%%%%%%%%%%%%%%%%%%%%%%%%%%%%%%%%%%%%%%%%%%%%%%%%%%%%%%%%%%%%%%%%%%%%%%%%

%%%%%%%%%%%%%%%%%%%%%%%%%%%%%%%%%%%%%%%%%%%%%%%%%%%%%%%%%%%%%%%%%%%%%%%%%%%%%%%%
\subsubsection{Robustness study over SNR}
Beyond the nominal-SNR configurations, Fig.~\ref{fig: snr_pd_pfa_iou_sim} evaluates the robustness of {\name} under varying SNR.
In simulation, SNR is defined as the ratio of signal to noise power over their respective occupied bandwidths for wideband sensing with heterogeneous signal bandwidths.
We sweep the SNR for representative scenarios in Figs.~\ref{fig: result_multi-config}\subref{subfig: sparse}--\subref{subfig: dense_ds3} and report the averaged $P_d$, $P_{fa}$, and IoU for $\thresIoU=0.5$.
As SNR increases, $P_d$, $P_{fa}$, and IoU exhibit overall upward trends due to improved signal visibility above the noise floor.
At low SNR, detections are dominated by spectrally compact signals (e.g., BLE), while signals with larger occupied bandwidths (e.g., OFDM/DSSS Wi-Fi) are more likely to fall below the detection threshold.
With increasing SNR, higher-bandwidth signals emerge more consistently, leading to higher $P_d$ and improved localization.
While ground-truth boundaries defined at fixed {3}\thinspace{dB} points remain unchanged, the effective detection boundaries expand with SNR, which can occasionally yield locally improved overlap at intermediate SNRs.
The behavior of $P_{fa}$ is likewise influenced by signal bandwidth and structure: at higher SNR, fragmented or partially detected higher-bandwidth signals become a more prominent source of false alarms.
Note that ground-truth blocks without any overlapping detection contribute zero IoU when averaging, which naturally bounds the averaged IoU by $\mathrm{IoU} \le P_d$.
Fragmented detections further compromise IoU, since each ground-truth block is matched to only its best detection, even when the overall energy is largely covered.

%%%%%%%%%%%%%%%%%%%%%%%%%%%%%%%%%%%%%%%%%%%%%%%%%%%%%%%%%%%%%%%%%%%%%%%%%%%%%%%%
%%%%%%%%%%%%%%%%%%%%%%%%%%%%%%%%%%%%%%%%%%%%%%%%%%%%%%%%%%%%%%%%%%%%%%%%%%%%%%%%
\subsection{Real-time Processing}\label{subsec: eval_rt}

We evaluate {\name}'s real-time performance to understand the computational bottlenecks and the requirements for meeting real-time deadlines in Sec.~\ref{subsec: rt_challenge} at {100}\thinspace{MHz} bandwidth.

\myparatight{Time breakdown.}
We profile the execution time of each image-processing stage in Sec.~\ref{subsec: imag_proc} using a single-threaded C++ implementation.
Fig.~\ref{fig: eval_time}\subref{subfig: time_breakdown} reports the average per-stage latency across the scenarios in Figs.~\ref{fig: result_multi-config}\subref{subfig: sparse}--\subref{subfig: dense_ds3} with $(T,F)=(2000,1024)$, with an average total latency of {339.75}\thinspace{\msec}.
Two-dimensional morphological operations (MO-2D) dominate the runtime, accounting for 67.58\% of the total execution time due to two-dimensional traversal and memory access patterns.

\myparatight{Multi-threaded with parallelism.}
To meet the real-time deadline, {\name} leverages parallel processing with multi-threaded execution (Sec.~\ref{subsec: sys_arch}).
We evaluate this design by running the representative workloads in Figs.~\ref{fig: result_multi-config}\subref{subfig: sparse}--\subref{subfig: dense_ds3} in a round-robin manner while varying the number of frequency partitions (i.e., threads).
With $(T,F)=(2000,1024)$ and $f_s={100}\thinspace\textrm{MSps}$, the processing deadline is {20.48}\thinspace{\msec}.
Fig.~\ref{fig: eval_time}\subref{subfig: ccdf_time_vs_subband} shows the latency CCDF over more than 1,000 TF plots varying the number of threads used.
{\name} achieves a 99th-percentile latency of {16.80}\thinspace{\msec} with 16 cores, satisfying the real-time requirement.
Doubling to 32 cores improves throughput but provides diminishing returns in tail latency.
Except for connected-component labeling (CCL), the execution time of {\name}'s processing stages is agnostic to the number of energy blocks.
The configuration in Fig.~\ref{fig: eval_time}\subref{subfig: ccdf_time_vs_subband} corresponds to  {100}\thinspace{MSps} raw I/Q input rate, and {\name} processes over 99\% of TF plots within the real-time deadline.

%%%%%%%%%%%%%%%%%%%%%%%%%%%%%%%%%%%%%%%%%%%%%%%%%%%%%%%%%%%%%%%%%%%%%%%%%%%%%%%%
%%%%%%%%%%%%%%%%%%%%%%%%%%%%%%%%%%%%%%%%%%%%%%%%%%%%%%%%%%%%%%%%%%%%%%%%%%%%%%%%
\subsection{Over-the-air (OTA) Evaluation}

Beyond simulation, we assess {\name} using OTA measurements for detection performance under realistic RF conditions.

%%%%%%%%%%%%%%%%%%%%%%%%%%%%%%%%%%%%%%%%%%%%%%%%%%%%%%%%%%%%%%%%%%%%%%%%%%%%%%%%

Fig.~\ref{fig: ota_controlled} evaluates {\name} under controlled OTA measurements using {100}\thinspace{MHz} bandwidth scenarios in Figs.~\ref{fig: result_multi-config}\subref{subfig: sparse}--\subref{subfig: dense_ds3}.
Fig.~\ref{fig: ota_controlled}\subref{subfig: snr_pd_ota} reports $P_d$ versus effective SNR at $\thresIoU=0.5$ for each scenario, where the averaged $P_d$ increases monotonically with SNR, consistent with simulation trends.
While absolute performance varies across scenarios, {\name} maintains reliable detection under realistic RF front-end effects.
The effective SNR is measured per scenario and averaged in the linear power domain rather than in dB.
Fig.~\ref{fig: ota_controlled}\subref{subfig: snr_pd_pfa_iou_ota} further summarizes the SNR dependence of average $P_d$, $P_{fa}$, and IoU.
Compared to simulation, OTA results exhibit a clearer trade-off between $P_d$ and $P_{fa}$, reflecting the influence of non-ideal RF front-end effects and non-stationary background noise that are not captured by idealized simulation.

%%%%%%%%%%%%%%%%%%%%%%%%%%%%%%%%%%%%%%%%%%%%%%%%%%%%%%%%%%%%%%%%%%%%%%%%%%%%%%%%
%%%%%%%%%%%%%%%%%%%%%%%%%%%%%%%%%%%%%%%%%%%%%%%%%%%%%%%%%%%%%%%%%%%%%%%%%%%%%%%%

\subsection{Baseline Comparison}

Fig.~\ref{fig: baseline_ota} compares Searchlight~\cite{bell2023searchlight}, DeepRadar~\cite{sarkar2021deepradar}, U-Net~\cite{uvaydov2024stitching}, and {\name} under a controlled OTA setting and summarizes their relative execution time, $P_d$, $P_{fa}$, and IoU across scenarios in Fig.~\ref{fig: result_multi-config}\subref{subfig: sparse}--\subref{subfig: dense_ds3}.
All methods are evaluated by offline replay of the same OTA captures at an average effective SNR of {14.08}\thinspace{dB}, as reported in Fig.~\ref{fig: ota_controlled}.
Under this OTA measurement, {\name} achieves an average IoU of 63.00\% and an average $P_d$ of 70.06\% under $\thresIoU=0.5$.

%%%%%%%%%%%%%%%%%%%%%%%%%%%%%%%%%%%%%%%%%%%%%%%%%%%%%%%%%%%%%%%%%%%%%%%%%%%%%%%%

\myparatight{Searchlight (convolution-based energy detection).}
Searchlight relies on thresholded convolution with kernel size search and power rate change to localize energy blocks.
To ensure a fair comparison, we tune its parameters on a representative scenario and apply the same configuration across all other cases.
Even under this tuned setting, Searchlight remains sensitive to signal density and time-frequency structure, resulting in fragmented detections and degraded performance, with an average $P_d$ of 35.79\% at $\thresIoU=0.5$ and an average IoU of 39.77\%.
In contrast, {\name} achieves higher $P_d$ and IoU with lower $P_{fa}$ while maintaining substantially lower processing latency.
By avoiding exhaustive kernel searches, {\name} achieves an average speedup of $20.51\times$ over Searchlight when both are implemented in Python, highlighting its algorithmic efficiency.

%%%%%%%%%%%%%%%%%%%%%%%%%%%%%%%%%%%%%%%%%%%%%%%%%%%%%%%%%%%%%%%%%%%%%%%%%%%%%%%%
\myparatight{DeepRadar (ML-based object detection).}
DeepRadar produces visually well-aligned detections with relatively low $P_{fa}$; however, its predicted bounding boxes often exhibit drift and size mismatches relative to the ground truth, leading to consistently low IoU.
As a result, DeepRadar's $P_d$ is particularly sensitive to the IoU threshold, where stricter $\thresIoU$ values lead to a sharp degradation in its detection performance.
Accordingly, we sweep $\thresIoU$ and report a representative operating point ($\thresIoU=0.2$) at which DeepRadar achieves usable detection performance, noting that lower IoU thresholds trade off spectrum localization accuracy for higher detection rates.
In addition, under OTA measurements, DeepRadar struggles to reliably detect certain signal types, indicating limited robustness for real-world applications.
We evaluate its PyTorch implementation against the single-core C++ version of {\name} on CPU, where {\name} achieves an average $1.65\times$ lower latency with a 56.02\% higher average IoU.
While DeepRadar can meet real-time constraints on GPUs, it exhibits limited robustness under OTA deployment.
In contrast, {\name} achieves a more favorable accuracy--latency trade-off on CPU.

%%%%%%%%%%%%%%%%%%%%%%%%%%%%%%%%%%%%%%%%%%%%%%%%%%%%%%%%%%%%%%%%%%%%%%%%%%%%%%%%
\myparatight{U-Net (ML-based image segmentation).}
Fig.~\ref{fig: baseline_ota} reports pixel-based $P_d$ and IoU for U-Net, with $P_{fa}$ not applicable, as discussed in Sec.~\ref{subsec: baseline_impl}.
Under OTA evaluation, U-Net achieves an average $P_d$ of 70.88\%, but exhibits a lower average IoU of 46.73\% and prohibitive computational overhead.
In Fig.~\ref{fig: baseline_ota}, the PyTorch-based U-Net on CPU is average {102.25}$\times$ slower than {\name} using a single CPU core, requiring average {34.55}\thinspace{s} per TF plot, with all methods evaluated on TF plots of identical pixel counts.
Even with GPU acceleration, U-Net requires {4.37}\thinspace{s} per TF plot, which is {213.38}$\times$ the real-time deadline, whereas {\name} meets the deadline with 16 CPU cores (Fig.~\ref{fig: eval_time}\subref{subfig: ccdf_time_vs_subband}).
Overall, the heavy architecture limits the practicality of U-Net for real-time wideband spectrum sensing.

%%%%%%%%%%%%%%%%%%%%%%%%%%%%%%%%%%%%%%%%%%%%%%%%%%%%%%%%%%%%%%%%%%%%%%%%%%%%%%%%

In summary, Fig.~\ref{fig: baseline_ota} highlights {\name}'s favorable accuracy-efficiency trade-off for real-time wideband spectrum sensing.

\section{Conclusion}
\label{sec: concl}

This paper presents {\name}, an image processing-based spectrum sensing system with real-time capability.
We formulate energy detection and localization as an image processing problem and analyze the effectiveness and computational complexity of the resulting techniques.
We define the real-time requirement and design a multi-threaded architecture to execute sensing tasks in parallel.
Detection metrics based on intersection over union (IoU) are defined, and {\name} is evaluated across multiple configurations, SNRs, and IoU thresholds.
Results show that {\name} meets the real-time processing requirement for {100}\thinspace{MHz} bandwidth sensing.
In controlled over-the-air experiments, {\name} achieves an average IoU of 62.22\% at an average effective SNR of {14.08}\thinspace{dB}, demonstrating robust detection under practical RF front-end effects.

\section*{Acknowledgments}
The work was supported in part by NSF under grants AST-2232458, ECCS-2434131, CNS-2443137, and CNS-2450567.

\bibliographystyle{ieeetr}
\bibliography{reference}

\end{document}